\documentclass[fleqn,usenatbib]{mnras}

\usepackage{newtxtext,newtxmath}  
\usepackage[T1]{fontenc}
\usepackage{ae,aecompl}

\usepackage{graphicx}	
\usepackage{amsmath}	
\usepackage{amssymb}	
\usepackage{bm}         
\usepackage{float}      
\usepackage{graphicx}   
\usepackage{url}        
\usepackage{booktabs}   
\usepackage{color}
\usepackage{hyperref}   
\usepackage[dvipsnames]{xcolor}     

\newcommand{\be}{\begin{equation}}
\newcommand{\ee}{\end{equation}}
\newcommand{\ba}{\begin{eqnarray}}
\newcommand{\ea}{\end{eqnarray}}

\newcommand*\mean[1]{\overline{#1}}
\newcommand*{\dt}[1]{\overset{\bm .}{#1}}

\definecolor{mygreen}{rgb}{0.19,0.55,0.11}
\definecolor{dkgreen}{rgb}{0,0.6,0}

\title[Modelling spin evolution of magnetars]{Modelling spin evolution of magnetars}

\author[Jawor \& Tauris]{Jędrzej A. Jawor$^{1}$\thanks{E-mail: jedrzejjawor@gmail.com},
Thomas M. Tauris$^{1}$\thanks{E-mail: tauris@phys.au.dk}
\\
$^{1}$Department of Physics and Astronomy, Aarhus University, Ny Munkegade 120, 8000~Aarhus~C, Denmark
}

\date{Accepted 2021 September 14. Received 2021 September 13; in original form 2021 June 1}

\begin{document}
\label{firstpage}
\pagerange{\pageref{firstpage}--\pageref{lastpage}}
\maketitle

\begin{abstract}
The origin and fate of magnetars (young, extremely magnetized neutron stars, NSs) remains unsolved. Probing their evolution is therefore crucial for investigating possible links to other species of isolated NSs, such as the X-ray dim NSs (XDINSs) and rotating radio transients (RRATs). Here we investigate the spin evolution of magnetars.  
Two avenues of evolution are considered: one with exponentially decaying B-fields, the other with sub- and super-exponential decay. 
Using Monte Carlo methods, we synthesize magnetar populations using different input distributions and physical parameters, such as for the initial spin period, its time derivative and the B-field decay timescale.  
Additionally, we introduce a fade-away procedure that can account for the fading of old magnetars, and we briefly discuss the effect of alignment of the B-field and spin axes.
Imposing the Galactic core-collapse supernova rate of $\sim 20\;{\rm kyr}^{-1}$ as a strict upper limit on the magnetar birthrate and comparing the synthetic populations to the observed one using both manual and automatic optimization algorithms for our input parameter study, we find that the B-field must decay exponentially or super-exponentially with a characteristic decay timescale of $0.5-10\;{\rm kyr}$ (with a best value of $\sim 4\;{\rm kyr}$). In addition, the initial spin period must be less than 2~sec. If these constraints are kept, we conclude that there are multiple choices of input physics that can reproduce the observed magnetar population reasonably well. 
We also conclude that magnetars may well be evolutionary linked to the population of XDINSs, whereas they are in general unlikely to evolve into RRATs.
\end{abstract}


\section{Introduction} \label{sec:intro}
Magnetars represent an extreme population of young neutron stars. Currently, about 30 magnetars are known in our Galaxy\footnote{McGill Online Magnetar Catalog \citep{ok14}: \url{http://www.physics.mcgill.ca/~pulsar/magnetar/main.html}}, observed either as soft gamma-ray repeaters (SGRs) or anomalous X-ray pulsars (AXPs).
Magnetars are characterized by having much larger spin periods ($1-10\;{\rm s}$) and larger spin period derivatives ($10^{-13}-10^{-11}\;{\rm s\,s}^{-1}$), compared to those of normal radio pulsars. As a result, their estimated surface B-fields ($B\,\propto\,\sqrt{P\dot{P}}$) are very large, typically $10^{14}-10^{15}\;{\rm G}$, whereas normal radio pulsars possess surface B-fields of $10^{11}-10^{13}\;{\rm G}$. 
These extremely strong magnetic fields are key in differentiating magnetars from other pulsars.  
Losses of magnetic energy are invoked in order to account for both their quiescent emission and their transient nature.
However, the exact way in which magnetic energy is converted into X-ray and $\gamma$-ray radiation is poorly understood.

Magnetars show magnetically driven enhancements of their thermal and non-thermal emission referred to as outbursts \citep{bl16,crp+18}. 
The transient phenomena include glitches, X-ray bursts and giant flares. 
Short bursts and flares, believed to involve the magnetosphere, are common observable magnetar activities, possibly triggered by their interior dynamics. For general reviews on magnetars, we refer to e.g. \citet{tzw15,kb17}.

Unlike radio pulsars, magnetars emit mainly high-energy X-ray and $\gamma$-ray radiation. Their quiescent emission usually consists of a softer, thermal component and a harder, non-thermal component that is believed to be produced in their magnetosphere \citep{tzw15}. 
Only five or six magnetars have been detected in the radio band. Their radio spectra are much flatter and thus assumed to be produced by a different mechanism than the one responsible for radio pulsar emission \citep{tzw15,kb17}. 
\citet{dvr+20} investigated their possible connection with fast radio bursts (FRBs) and 
at least in one case, a magnetar has been associated with a FRB \citep[the Galactic SGR~1935+2154,][]{brb+20,ksj+20,rsf+21}.

\begin{figure*}
\centering
\vspace{-1.5cm}
\includegraphics[width = 1.10\textwidth, angle=0]{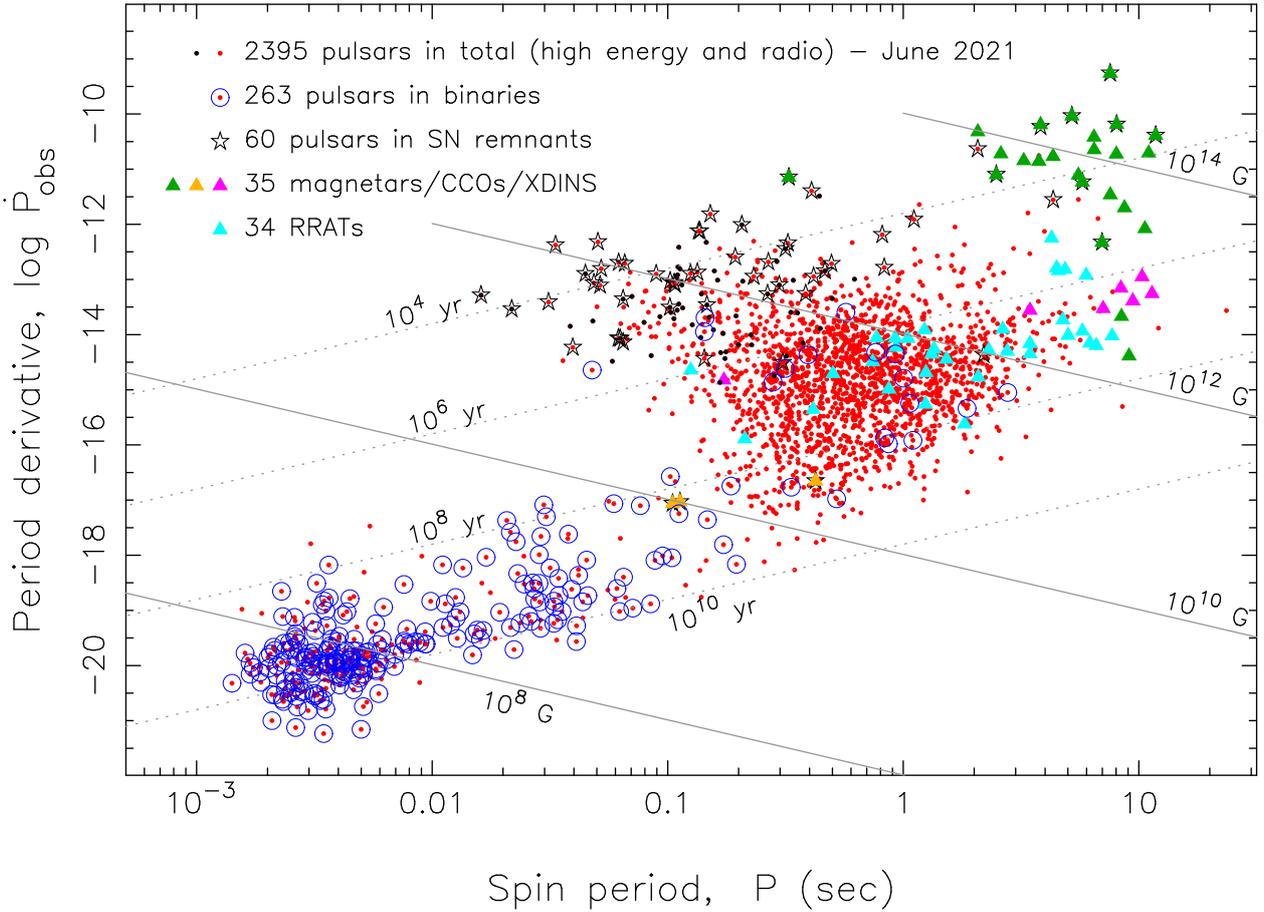}
\vspace{-1.5cm}
\caption[PdP]{$P\dot{P}$--diagram. Red dots represent radio pulsars. Triangles in various colours represent magnetars, core compact objects (CCOs), X-ray dim isolated neutron stars (XDINSs) and rotating radio transients (RRATs).
Objects associated with supernova remnants or binary systems are overplotted with stars and circles respectively.
Also plotted are lines of constant B-field and spin-down age as derived from the dipole model presented in Section~\ref{sec:psr-evol}.
After \citet{tv22}. Data is from the ATNF catalogue in June~2021: \url{http://www.atnf.csiro.au/research/pulsar/psrcat} \citep{mhth05}.}
\label{fig:PdP_example}
\end{figure*}

The spin evolution of magnetars is closely related to that of radio pulsars in view of the magnetodipole model \citep{pac67,mt77,st83,lk04}.
To this day, more than 3000 radio pulsars have been detected\footnote{ATNF Pulsar Catalogue \citep{mhth05}: \url{https://www.atnf.csiro.au/research/pulsar/psrcat/}.}, see the $P\dot{P}$--diagram for all neutron stars (NSs) in Fig.~\ref{fig:PdP_example}.
As can be seen from this plot, the known population of NSs is mainly separated into three distinct types: normal radio pulsars (the central bulk of the population), recycled (rapidly spinning and low B-field) millisecond pulsars (MSPs) in the lower left corner, and magnetars in the upper right corner. As indicated in the figure, magnetars are often associated with supernova remnants, which is direct evidence for their young nature.

Unquestionably, the different types of NSs are quite diverse. The question is then: {\em How are these objects connected to each other?}
The ``Grand Unification of Neutron Stars'' (GUNS) model proposes that rotating and magnetized NSs are manifestations of the same underlying physics \citep{kas10}. In this picture, the differences between the distinct types are mainly caused by discrepancies in age and initial magnitude of the B-field. 
The latter may depend on the degree of supernova (SN) fallback \citep{zkst21}.
Hence, the main difference between a radio pulsar and a magnetar is believed to originate from their initial B-field strength. The B-field of a normal radio pulsar simply lacks the power to initiate the processes which produce the X-ray emission and bursts of magnetars. 
In the GUNS picture, it is possible that magnetars evolve into other types of NSs if their B-fields decay over time \citep{vrp+13}. 
Such evolutionary links are also a possible solution to the ``NS birth rate problem'' \citep{kk08}.

In this paper, we investigate the spin evolution of rotating NSs. The focus will be on the evolution of magnetars; however, the derived spin evolution equations could be applied to any type of rotating and magnetized NS. 
Various models will be derived in order to determine how the B-fields of magnetars decay with time and we briefly investigate the possibility of magnetars evolving into the so-called X-ray dim NSs (XDINSs) or becoming rotating ratio transients (RRATs), see e.g. \citet{rip+13,kk16,kb17}. The applied method will be a combination of analytical and numerical investigations as well as a population synthesis study with a comparison to observations.

In Section~\ref{sec:psr-evol}, we consider the dipole braking model which is assumed to account for the spin-down of magnetars. 
In Section~\ref{sec:synth-pop}, two evolutionary avenues are defined 
and applied to produce synthetic populations of magnetars (using Monte Carlo methods) which are compared to the observed magnetars 
in the search for an optimal set of parameters describing their evolution. A fade-away procedure is introduced to account for older magnetars fading from detection. 
In Section~\ref{sec:optimizing}, we analyse and optimize the results from Sections~\ref{sec:psr-evol} and \ref{sec:synth-pop}. 
A goodness of fit is used together with the synthetic birth rate to constrain the ranges of free parameters that can account for the observed magnetar population. Afterwards, the differences between different viable models are discussed. 
Further discussions on optimizing algorithms, best parameter search, initial parameter distributions and fade-away are given in Section~\ref{sec:discussion}, including the question of linking faded magnetars to the populations of XDINSs or RRATs.
Finally, in Section~\ref{sec:summary} we summarize our findings and briefly discuss directions for future work.

\section{Pulsar evolution}\label{sec:psr-evol}
Knowledge of spin evolution is the first step in determining what kinds of models best describe the observed magnetars, as it allows for direct comparison with the observed sample in the $P\dot{P}$--diagram. 
In the following, we briefly summarize the braking mechanism that causes the spin-down of radio pulsars and demonstrate how it depends on the B-field and inclination angle between the magnetic and spin axes. 
We then introduce functions describing the temporal evolution of the B-field and describe two avenues of spin evolution.

\subsection{Braking mechanisms}
In short, an accelerated magnetic dipole leads to emission of electromagnetic (magnetodipole) waves with a frequency equivalent to the spin frequency of the pulsar. The required energy for this emission is tapped from the rotational energy reservoir of the NS and is responsible for producing a braking torque:
\begin{equation}\label{eq:braking_torque}
    N \equiv \frac{d J}{d t} = I \dot{\Omega} \,,
\end{equation}
where $J$ is the spin angular momentum of the NS, $J = I \Omega$, $I$ is the moment of inertia (assumed to be constant), and $\Omega = 2\pi/P$ is its angular spin velocity. For all isolated and non-accreting pulsars $\dot{\Omega}<0$, i.e. the pulsars slow down over time.
In order to study the evolution of pulsars in the $P\dot{P}$--diagram the time dependence of $N$ must be found \citep[e.g.][]{tk01}.

\subsubsection{The dipole model}
In the dipole model, it is assumed that the surface B-field of the pulsar is a pure dipole \citep{pac67,mt77}.
The dipole model is an {\em in vacuo}, oblique rotator model, i.e. it assumes that there is vacuum right outside the surface of the pulsar and requires a misalignment between the rotation and B-field axes.
The B-fields of magnetars will probably have toroidal and poloidal components and thus deviate from the dipole form \citep{vrp+13,tzw15,pv19}. However, a pure dipole is still a good starting point, as the magnitude of the toroidal component and its influence on spin-down are hard to estimate.
In addition, it is assumed that only the surface B-field, also called the crustal field, is responsible for generating the braking torque. Therefore, the interior of the NS plays, in principle, no direct role. This is an advantage, as it makes it possible to evaluate the instantaneous spin-down torque without taking into account the complex interior. 
The energy-loss rate due to magnetic dipole radiation is given by:
\begin{equation}
  \dot{E}_{\rm dipole} = -\frac{2}{3c^3}|\ddot{\bm m}|^2\,,
  \label{eq:Edipole}
\end{equation}
where $|\ddot{\bm m}|=BR^3\Omega^2\sin\alpha$ is the second time derivative of the magnetic moment of the NS, $B$ is the magnetic flux density at its surface (equator), 
$R$ is the radius,
$\Omega = 2\,\pi /P$ is the angular velocity with $P$ being the pulsar spin period,
$c$ is the speed of light in vacuum, and the magnetic inclination angle is $0<\alpha\le \pi/2$.

\subsubsection{Magnetospheres}
The dipole model neglects the existence of a plasma-filled magnetosphere, which is not a good assumption since electromagnetic forces are able to rip particles off the NS surface, not to mention its necessity for explaining the emission mechanism of radio pulsars or the burst activities of magnetars. 
Of particular interest here is that the existence of a magnetosphere can contribute significantly to the spin-down of a pulsar. This is usually attributed to currents permeating the magnetosphere (giving rise to the $\vec{j}\times\vec{B}$ force exerted by the plasma current) and outflows of plasma-loaded winds beyond the light cylinder \citep{gj69,mic82,cor90,spi06}.

Despite the evidence that magnetospheres play some role in spin evolution, it is uncertain exactly how important they are \citep{vrp+13,gmvp14,ton16,kb17}. The magnitude of the braking torque from most magnetospheric models is quite similar to that of the vacuum dipole model \citep[e.g.][]{st83,spi06} and both produce default evolution with a braking index of $n=3$.
Furthermore, observations of radio pulsars have shown that $\dot{E}_{\rm plasma}$ is of the same order as $\dot{E}_{\rm dipole}$ within a factor of a few \citep{klo+06,llm+12,crc+12}.
For these reasons, we choose to only consider the dipole term for the braking torque magnitude.
Finally, we assume that the mechanisms responsible for the persistent X-ray emission and transient phenomena of magnetars do not contribute to their overall braking torque.

\subsubsection{The braking law}
The energy source that powers the dipole radiation originates from the rotational energy of the pulsar:
\begin{equation}
  E_{\rm rot} = \frac{1}{2} I \Omega^2 \,.
\end{equation}
Equating the loss rate of rotational energy to the emitted dipole power, $\dot{E}_{\rm rot} = \dot{E}_{\rm dipole}$ yields:
\begin{equation}
  \dot{\Omega} = -\frac{2 R^6}{3 Ic^3}\,B^2 \Omega^3 \sin^2 \alpha \,.
\label{eq:Dip_EroteqEdip}
\end{equation}
In general, the braking index, $n$, of the spin deceleration of a pulsar is defined by \citep{mt77}:
\begin{equation} 
\label{eq:n}
    \dot{\Omega} \propto -\Omega ^n \,,
\end{equation}
which yields (for $n$ constant): $n\equiv \Omega\ddot{\Omega}/\dot{\Omega}^2$.
This deceleration law can also be expressed as: 
$\dot{P} \propto P^{2-n}$ and hence the
slope of a pulsar evolutionary track in the $P\dot{P}$--diagram is simply given by: $2-n$.
Depending on the physical conditions under which the pulsar spins down, $n$ can take different values.
For example: $n>3$ for pulsars with B-field decay, multipoles, or alignment (Section~\ref{subsubsec:alignment}).
The combined magnetic dipole and plasma current spin-down torque may also result in $n \ne 3$ \citep{cs06}.
A positive braking index, $n > 0$, means that $\ddot{\Omega}>0$ and thus the magnitude of the braking torque, $|N|$ is decreasing over time.

A simple integration of equation~(\ref{eq:n}) for a {\em constant} braking index ($n\ne1$)
yields the well-known expression:
\begin{equation} 
\label{eq:trueage}
   t=\frac{P}{(n-1)\dot{P}}\left[ 1-\left(\frac{P_0}{P}\right)^{n-1}\right] \,,
\end{equation}
where $t$ is the so-called true age of a pulsar, which had an initial spin period $P_0$ at time $t=0$. 
The characteristic age (or the spin-down age), $\tau$ is defined as the true age for a pulsar with $n=3$ and $P_0\ll P$, i.e. $\tau\equiv P/2\dot{P}$.

\subsubsection{Beyond the dipole model}\label{subsubsec:eta}
A more general form of equation~(\ref{eq:Dip_EroteqEdip}) is given by \citep[e.g.][]{bhvk19}:
\begin{equation}
  \dot{\Omega} = -\frac{2 R^6}{3 Ic^3}\,B^2 \Omega^\eta \sin^2 \alpha \,,
\label{eq:Dip_EroteqEdip_w_eta}
\end{equation}
where the $\eta$ parameter introduced here can be used to explore spin evolution beyond the dipole model. $\eta$ mainly affects evolution before the decay of the B-field and/or $\alpha$, as in this early regime $\dot P \propto P^{2-\eta}$ for constant $B$ and $\alpha$. We produced models with different values of $\eta$, but found that they are not better at reproducing the observed population of magnetars. As there is no reason to introduce additional degrees of freedom, we choose to only work with the traditional dipole formulation, i.e. $\eta = 3$. 
However, see Appendix~\ref{AppendixE}.

\subsection{Temporal evolution of spin, B-field and inclination angle}
Rewriting equation~(\ref{eq:Dip_EroteqEdip}) and introducing $K\equiv 8R^6\pi^2/3Ic^3$, such that in terms of the spin period, $P$ as a function of time, $t$:
\begin{equation}
\label{eq:Pdot}
    \dot{P}(t) = K B(t)^2 \sin^2\alpha(t) \ P(t)^{-1} 
\end{equation}
For the constant, we chose $K=8.77\times 10^{-40}\;{\rm cm\,s^3\,g^{-1}}$ as our default value. In Section~\ref{subsubsec:EoS}, we briefly discuss the effect of using other values of K (i.e. changing the NS equation-of-state). 
In order to evaluate the effects of a time dependent B-field and magnetic inclination angle on spin-down, we need to establish a model for how they evolve with age, i.e. we need to find $B(t)$ and  $\sin\alpha(t)$.

\subsubsection{B-field decay}
In general, the evolution of the B-field is connected to the somewhat poorly known equation-of-state and cooling models of NSs. 
We model the decay of the B-field using the analytical expression \citep{cgp00,bhvk19}:
\begin{equation}
  \frac{d B(t)}{d t} =  -a B(t)^{1+\beta} \,,
\end{equation}
where $a$ and $\beta$ are model parameters.
This expression allows us to approximate the results from works that solve combined thermal and magnetic evolution \citep[e.g.][]{gu94,tk01,pmg09,vrp+13} using numerical means. Such an approach is beyond the scope of this paper. 
Solving the above equation yields:
\begin{align}\label{eq:B}
  &B(t) = 
  \begin{cases}
  B_0 \left(1+ \displaystyle\frac{\beta\,t}{\tau_B} \right)^{\frac{-1}{\beta}}  & \beta \neq 0\,, \\
  B_0 \,e^{-t/\tau_B}  & \beta = 0\,.
  \end{cases}
\end{align}
Here, $\tau_B = B_0^{-\beta}/a$ is the characteristic decay timescale for the B-field and $B_0$ is the initial value of the surface B-field.
The $\beta$ value controls how the B-field decays: $\beta < 0$ corresponds to super-exponential decay, $\beta > 0$ to sub-exponential decay. 
The special case of $\beta=0$ corresponds to the classic exponential decay obtained by considering a pulsar with a B-field confined to the crustal regions and decaying due to diffusion and Ohmic dissipation \citep{gu94,tk01}.

\subsubsection{Alignment}\label{subsubsec:alignment}
There is evidence from observations of radio pulsars that the magnetic field axis aligns with the spin axis on a long timescale \citep{tm98,jk17}.
Here we follow \citet{jon76,tk01} who applied a simple exponential expression for $\alpha(t)$:
\begin{equation}\label{eq:inclination_exp}
  \sin\alpha(t) = \sin\alpha_0\,e^{-t/\tau_\alpha} \,,
\end{equation}
where $\tau_\alpha$ is the characteristic decay timescale of the inclination angle and $\alpha_0$ is the initial inclination angle. 
We disregard here the treatment of alignment as a separate effect. Instead we may consider a combined timescale from exponential B-field decay and exponential alignment by considering a single effective reduced timescale:  $\tilde{\tau} = \tau_B \tau_\alpha/(\tau_B + \tau_\alpha)$ \citep{tk01}.
For magnetars, a similar alignment has been suggested on a much shorter timescale of a few hundred years \citep{lj20}.

\subsection{Evolution in the $P\dot{P}$--diagram}\label{subsec:Pevolve}
Applying equation~(\ref{eq:B}) for $B(t)$ and setting $\alpha(t)=\alpha_0$, we can integrate equation~(\ref{eq:Pdot}) to find $P(t)$:
\begin{align}
   P(t) =
   \begin{cases}
   \sqrt{ \displaystyle\frac{2 K B_0^2 \, \sin^2\!\alpha_0 \, \tau_B}{\beta-2}\,\left[\left( 1+\frac{\beta\, t}{\tau_B}\right) ^{\frac{\beta-2}{\beta}} -1\right] + P_0^2  } & \beta \neq 0 \,, \vspace{0.2cm} \\ 
    \sqrt{K B_0^2 \, \sin^2\!\alpha_0 \, \tau_B \left(1 -\exp \left(\displaystyle\frac{-2t}{\tau_{B}}\right)\right) + P_0^{2}} & \beta = 0\,.
\end{cases}
\label{eq:Both_avenues_P}
\end{align}
Knowing $P(t)$, equation~(\ref{eq:Pdot}) can be used again to find $\dot{P}(t)$. Knowing these two values, allows for the full description of the spin evolution of pulsars and production of evolutionary tracks in the $P\dot{P}$--diagram \citep{tk01}. It also makes it possible to find the braking index, $n$ and the true age, $t$ from equation~(\ref{eq:trueage}).
Most importantly, the modelling of trial evolutionary tracks can be used to draw some rough conclusions on which values of parameters ($B_0$, $\tau _B$, $\beta$, $\alpha_0$), if any, can reproduce the distribution of the observed magnetar population in the $P\dot{P}$--diagram.

\begin{figure*}
\centering
\includegraphics[width = \textwidth]{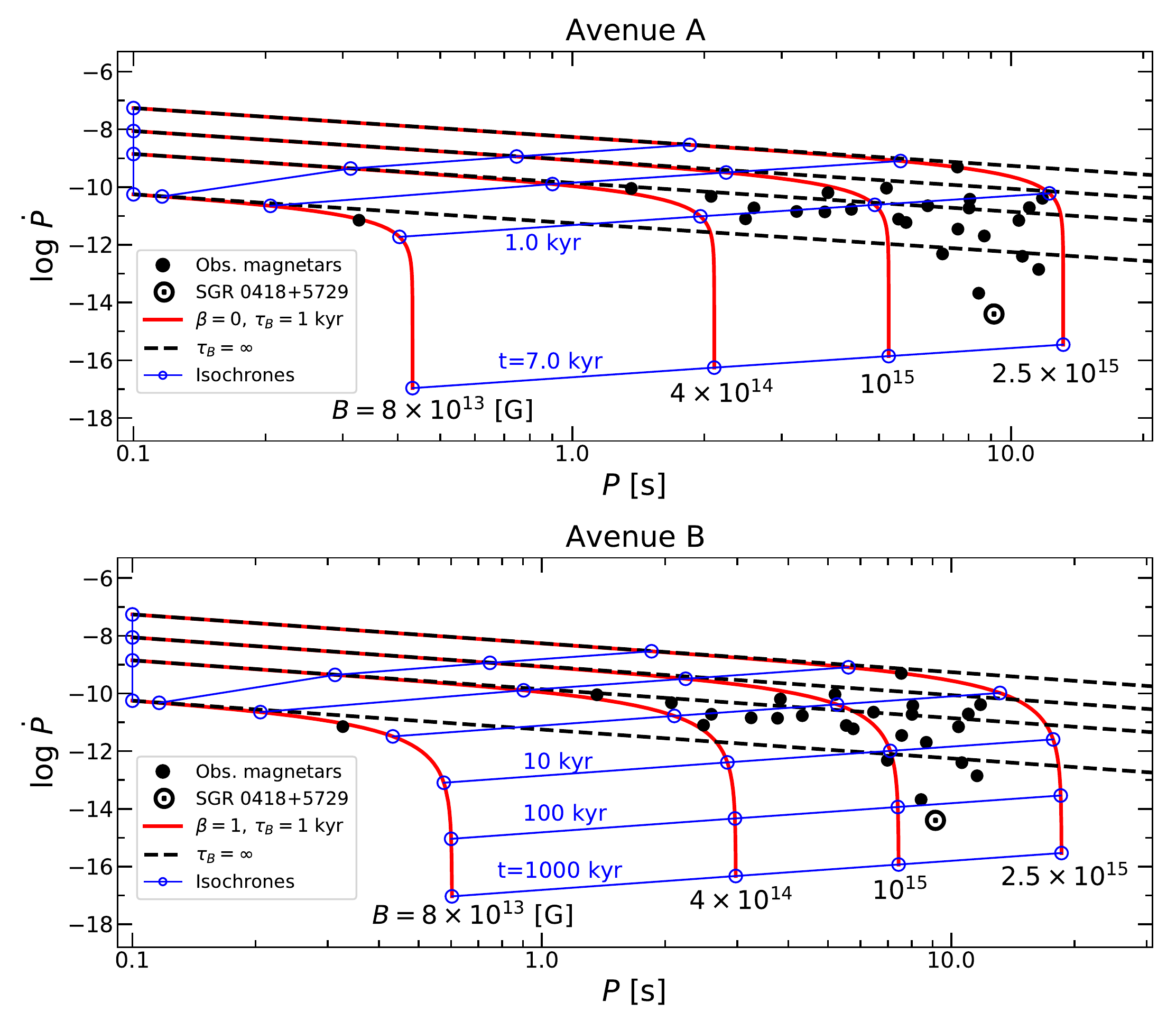}
\caption{Evolutionary tracks from Avenue~A (top) and Avenue~B (bottom). The tracks with decaying B-fields  (red lines) are calculated using $\tau _B=1\;{\rm kyr}$, different values of $B_0$ (see labels) and $P_0 = 0.1\;{\rm s}$. The solid black dots represent observed magnetars \citep{ok14}. The blue lines and circle are isochrones which mark the true ages. Top: $t = \{0,\,0.01,\,0.1,\,1.0,\,7.0\;{\rm kyr}\}$. Bottom: $t = \{0,\,0.01,\,0.1,\,1.0,\,10,\,100,\,1000\;{\rm kyr}\}$. The black, dashed lines are evolutionary tracks in the absence of B-field decay and alignment.  
SGR~0418+5729 \citep{rip+13} is shown with a circle.}
\label{fig:Evo_tracks_both_avenues}
\end{figure*}

We define two evolutionary avenues, A and B. Avenue~A corresponds to the $\beta = 0$ case (exponential decay), while Avenue~B is the $\beta \neq 0$ case. 
Evolutionary tracks from both avenues are plotted in Fig.~\ref{fig:Evo_tracks_both_avenues}. 
In both panels, one set of tracks is calculated with $\tau_{B} = 1\;{\rm kyr}$ and another with $\tau_B = \infty$ (i.e. constant B-field). Both avenues assume: an initial spin period, $P_0 = 0.1\;{\rm s}$, a fixed inclination angle $\alpha_0=\pi/2$, and initial B-fields of $B_0 = \{8.0\times 10^{13}, \,4\times 10^{14}, \,1.0\times 10^{15}, \,2.5\times 10^{15}\;{\rm G}\}$. These $B_0$ values are chosen such that the tracks intersect the region with the observed magnetars. 
At early ages, the two sets of tracks for each avenue are identical. The sets with finite $\tau_{B}$ start to bend down when $t\approx\tau_{B}$ since the braking torque weakens together with the B-field causing the spin period to approach an asymptotic value.

Comparing the two different avenues, in the case of the sample tracks shown in Fig.~\ref{fig:Evo_tracks_both_avenues}, the evolution is similar up to about $t \simeq 1\;{\rm kyr}$. In Avenue~A (exponential decay), however, the further evolution is significantly accelerated compared to that of Avenue~B.
Thus, for the specific Avenue~A tracks presented here, the true age of the oldest observed magnetar \citep[SGR~0418+5729,][]{rip+13} is only about 5~kyr, whereas it is above 100~kyr according to the specific Avenue~B tracks. Notice that these quoted ages are only valid for one selected set of models. In general, 
the difference between models of Avenues~A and B may vary less or even more, depending on the adopted value of $\beta$ and $\tau _B$.

\begin{figure*}
\centering
\includegraphics[width = \textwidth]{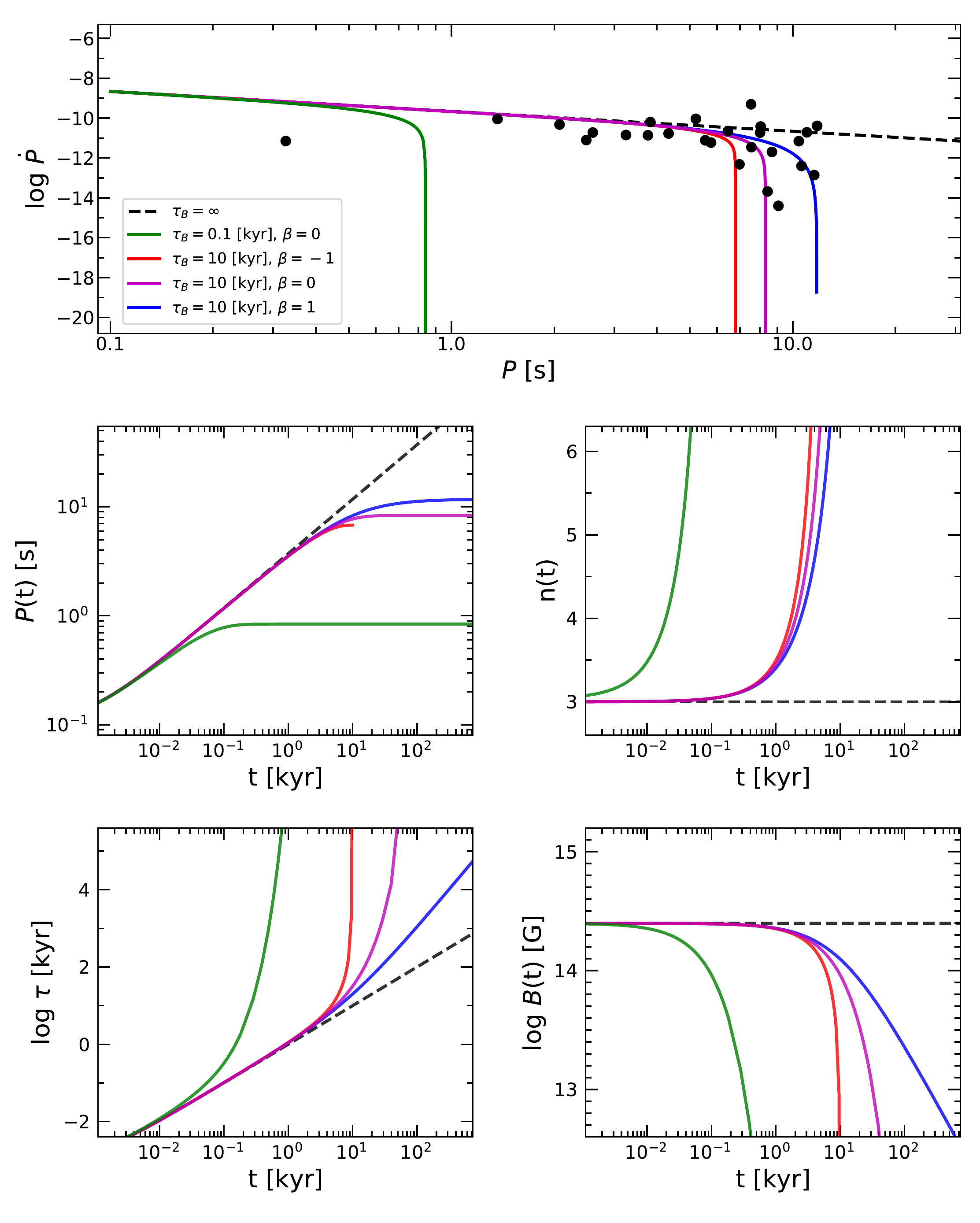}
\caption{Evolution of five different models: two from each of Avenues~A and B, and one model with constant B-field ($\tau_B=\infty$).
Top: $P\dot{P}$--diagram, center left: spin period, center right: braking index, bottom left: spin-down age, bottom right: B-field. The dots represent observed magnetars \citep{ok14}.
The various models all have initial values: $P_0 = 0.1\;{\rm s}$, $B_0 = 5.0\times 10^{14}\;{\rm G}$, $\alpha_0=\pi/2$, but different types of B-field decay determined by $\tau_B$ and $\beta$, see legend.}
\label{fig:Params_both_avenues}
\end{figure*}

Figure~\ref{fig:Params_both_avenues} illustrates how changes in parameters affect the spin evolution. We have plotted: spin periods, braking indices, spin-down ages, and B-fields, as functions of true age for five different spin evolution models calculated with $\tau_{B} = \{0.1\;{\rm kyr}, \,10\;{\rm kyr}, \,\infty \}$ and three different choices of $\beta$ for $\tau_B=10\;{\rm kyr}$. Two of the models follow Avenue~A, two follow Avenue~B, while the last model has a constant B-field (i.e. $\beta =0$ and $\tau_B =\infty$).

As already shown in Fig.~\ref{fig:Evo_tracks_both_avenues}, $B_0$ affects the initial $\dot{P}$ value ($\dot{P}_0$) of the evolutionary tracks. Together with $P_0$ and $\alpha_0$, it completely specifies the starting point of an evolutionary track. 
$\tau_{B}$ affects the location in the $P\dot{P}  $--diagram where the evolutionary track bends down. Looking at the bottom right panel of Fig.~\ref{fig:Params_both_avenues}, when the true age approaches $\tau_{B}$, the B-field starts to decay significantly. 
When this happens, the braking torque decreases causing the spin period to converge to a constant value. Thus, from equation~(\ref{eq:braking_torque}) it is clear that $|\dot{\Omega}|\rightarrow 0$ and therefore, by definition, the braking index, $n\rightarrow \infty$, as displayed in the center right panel of Fig.~\ref{fig:Params_both_avenues}.
The case with $\tau_{B} = \infty$ has a constant $n=3$ and $B(t)=B_0$, meaning that $\dot{P}\,\propto\,P^{-1}$ at all ages. 
Figure~\ref{fig:Params_both_avenues} also shows the spin-down (characteristic) age as function of true age. For the model with $\tau_{B}=\infty$ , the two ages are identical (except for a very young pulsar where $P\simeq P_0$). In contrast, for the two other models, $\tau$ diverges once $t$ approaches (and grows beyond) $\tau_{B}$. This result demonstrates why the spin-down age is a bad measure of the true age of a pulsar if its B-field has decayed.

$\beta$ directly affects the evolution of the B-field. Choosing $\beta<0$ causes the B-field to decay super-exponentially, whereas $\beta>0$ causes the B-field to decay sub-exponentially. Faster decay translates into a sharper bend of the evolution track, while slower decay makes the bend smoother.
The model with $\beta = -1$ actually reaches $B(t) = 0$ at a finite age. This age can be found by solving for it in equation~(\ref{eq:B})\footnote{This is not a physical result and a realistic NS may have an additional, non-decaying component of the B-field, e.g. anchored in its core.}:
\begin{equation}
  t(B=0) =  -\frac{\tau_{B}}{\beta}, \quad \beta < 0.
\label{eq:Ave_C_negative_beta_zero_time}
\end{equation}
At this age, the braking index becomes infinite and the spin period constant. This behaviour is true for all models with $\beta < 0$.
Note that in the model with $\beta = -1$ plotted in Fig.~\ref{fig:Params_both_avenues}, $P(t)$ is cut off exactly when $B(t)$ becomes 0, which is why it stops before all other models.

\section{Synthetic population of magnetars}\label{sec:synth-pop}
Studying evolutionary tracks alone does not provide good answers to which avenues and parameters best describe the evolution of real magnetars. In order to answer this question, we generate synthetic populations and evolve them using the spin evolution equations from the previous section. In addition, we introduce a fade-away procedure that can account for the synthetic magnetars fading from detectability over time.
Finally, we apply statistical tests in order to compare synthetic populations to the observed one in the $P\dot{P}$--diagram. The results of these tests are used to optimize parameters in search for the best fitting models.

\subsection{Observed population of magnetars}\label{subsec:ObsSample}
The properties of the observed magnetars are taken from the McGill catalogue \citep{ok14}\footnote{The catalogue can be found at: \url{http://www.physics.mcgill.ca/~pulsar/magnetar/main.html}}.
This updated catalogue contains information on 31 magnetars in total, six of which are unconfirmed. In addition, the catalogue includes the pulsar PSR~J1846$-$0258 ($P=0.324\;{\rm s}$) located in the supernova remnant Kes~75. This NS was initially classified as a young rotation-powered pulsar; however, in 2006 and 2020, it underwent magnetar-like outbursts \citep{ggg+08,bsmf21}. Consequently, PSR~J1846$-$0258 is sometimes classified as a magnetar and is included as such in the subsequent analysis.

\begin{figure}
\centering
\includegraphics[width = \columnwidth]{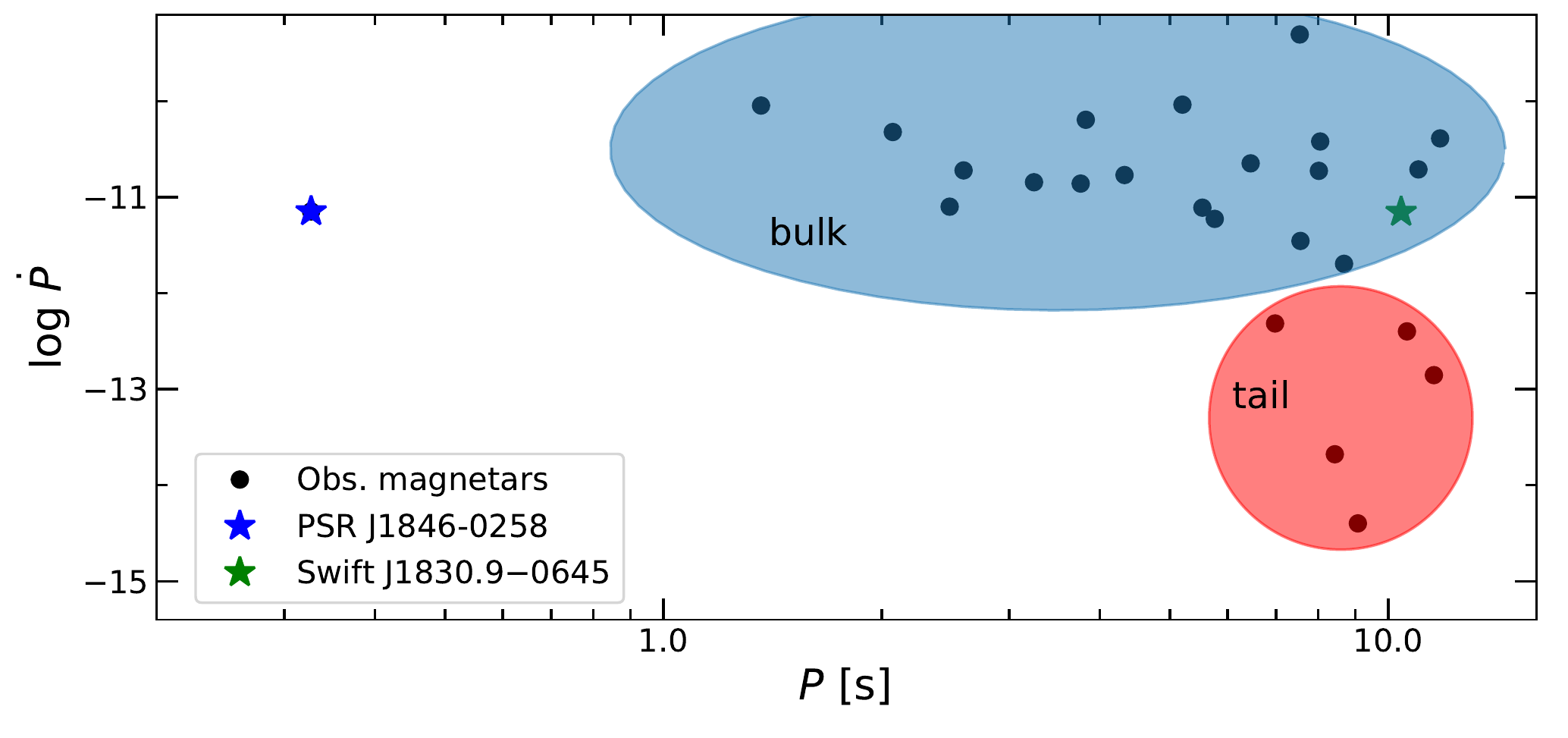}
\caption{$P\dot{P}$--diagram with the observed magnetars. The outlier PSR~J1846$-$0258 and the new magnetar Swift~J1830.9$-$0645 are marked with stars. The blue and red shaded areas enclose the ``bulk'' and the ``tail'' of the observed population, respectively.}
\label{fig:ObsMag}
\end{figure}

Our ensemble of observed magnetars consists of all confirmed magnetars with known $P$ and $\dot{P}$ from the McGill catalogue and the newly discovered Swift~J1830.9$-$0645 \citep{cbi+20}, 26~objects in total. The values of $P$, $\dot{P}$ and $B$ of this ensemble are listed in Table~\ref{table:ObsMag} and a $P\dot{P}$--diagram is plotted in Fig.~\ref{fig:ObsMag}.
From this plot, we see that PSR~J1846$-$0258 is clearly an outlier in the $P\dot{P}$--plane as it has a much smaller $P$ compared to the other magnetars. This separation from the rest of the magnetar population may be a sign of a distinct evolutionary path --- possibly related to this magnetar being the only one observed as a radio pulsar.

\begin{table}
\caption[]{Spin properties and B-fields of all observed magnetars. The B-fields are estimated in the dipole model: $B_{\rm obs} = 3.2 \times 10^{19} \sqrt{P\dot{P}}\;{\rm G}$. References can be found in the McGill catalogue \citep{ok14}.}
\begin{tabular}{lrcc}
\hline
  Name & $P$              & $\log(\dot{P})$           & $\log(B_{\rm obs})$ \\
       & (s)              & ({\rm s\,s}$^{-1}$)       & (G) \\
\hline
\noalign{\smallskip}
PSR J1846$-$0258          & 0.33 & $-$11.15 & 13.69 \\
Swift J1818.0$-$1607      & 1.36 & $-$10.04 & 14.55 \\
1E 1547.0$-$5408          & 2.07 & $-$10.32 & 14.50 \\
Swift J1834.9$-$40846     & 2.48 & $-$11.10 & 14.15 \\
SGR 1627$-$41             & 2.60 & $-$10.72 & 14.35 \\
SGR 1935+2154             & 3.25 & $-$10.84 & 14.32 \\
SGR J1745$-$2900          & 3.76 & $-$10.86 & 14.36 \\
CXOU J171405.7$-$381031   & 3.83 & $-$10.19 & 14.70 \\
PSR J1622$-$4950          & 4.33 & $-$10.77 & 14.44 \\
SGR 1900+14               & 5.20 & $-$10.04 & 14.85 \\
XTE J1810$-$197           & 5.54 & $-$11.11 & 14.32 \\
SGR 0501+4516             & 5.76 & $-$11.23 & 14.27 \\
1E 1048.1$-$5937          & 6.46 & $-$10.65 & 14.59 \\
1E 2259+586               & 6.98 & $-$12.32 & 13.77 \\
SGR 1806$-$20             & 7.55 & $-$9.31 & 15.29 \\
SGR 1833$-$0832           & 7.57 & $-$11.46 & 14.22 \\
CXOU J010043.1$-$721134   & 8.02 & $-$10.73 & 14.59 \\
SGR 0526$-$66             & 8.05 & $-$10.42 & 14.75 \\
Swift J1822.3$-$1606      & 8.44 & $-$13.68 & 13.13 \\
4U 0142+61                & 8.69 & $-$11.69 & 14.13 \\
SGR 0418+5729             & 9.08 & $-$14.40 & 12.79 \\
Swift J1830.9$-$0645      & 10.42& $-$11.15 & 14.74 \\
CXOU J164710.2$-$455216   &10.61 & $-$12.40 & 13.82 \\
1RXS J170849.0$-$400910   &11.01 & $-$10.71 & 14.67 \\
3XMM J185246.6+003317     &11.56 & $-$12.85 & 13.61 \\
1E 1841$-$045             &11.79 & $-$10.39 & 14.85 \\
\hline
\end{tabular}
\label{table:ObsMag}
\end{table}

The distribution of observed magnetars can be seen to bend downwards with increasing $P$, never crossing $P=12\;{\rm s}$. We separate it into two sub-populations. The {\em bulk} contains 20 of the magnetars which lie between $\dot{P} =10^{-9.30}$ and $\dot{P}=10^{-12}\;{\rm s\,s}^{-1}$. Meanwhile, the {\em tail} consists of the remaining five magnetars with $\dot{P}$ below $10^{-12}\;{\rm s\,s}^{-1}$.
Both sub-populations are marked in Fig.~\ref{fig:ObsMag} by shaded areas.

The reason for this split is based on the forms of evolutionary tracks found in Section~\ref{sec:psr-evol}. If the B-field decays, then the magnetars in the tail and bulk could be explained by similar tracks. However, if $B$ is constant, then the tracks passing through the tail magnetars will be different from the bulk ones. 
This distinction will prove useful when dealing with synthetic populations.

\subsection{Zero-age population of synthetic magnetars}\label{subsec:ZeroAge}

\begin{table}
\centering
\begin{tabular}{ccc}
\toprule
Parameter & Distribution function & Variables\\
\midrule
$P_0$& log-normal~\citep{jkb95}& $\mu_{P_0}$, $\sigma_{P_0}$ \\
$\dt{P}_0$ & log-normal~\citep{jkb95} & $\mu_{\dt{P}_0}$, $\sigma_{\dt{P}_0}$ \\
$\alpha_0$ & $\sin \alpha_0$ & --- \\
$t$ & uniform & $t_{\rm max}$ \\
\bottomrule
\end{tabular}
\caption{Distribution function used to generate the synthetic zero-age populations.}
\label{tab:zero_age}
\end{table}

A zero-age population of $N$ synthetic newborn magnetars is defined with initial spin periods, $P_0$, spin period derivatives, $\dot{P}_0$, and magnetic inclination angles, $\alpha_0$ (Table~\ref{tab:zero_age}). 
For $P_0$ and $\dt{P}_0$, $\mu$ is the expectation value and $\sigma$ is the standard deviation of the natural logarithm. 
In the models shown in Tables~\ref{tab:manualmodels} and \ref{tab:automodels}, we assume a value between $1.6-16\;{\rm ms}$ for $P_0$ drawn from a log-normal distribution with $\mu_{P_0}=0.005\;\rm{s}$ and $\sigma_{P_0}=0.5$. (The lower and upper limits of 1.6~ms and 16~ms simply reflect the 99\% level boundaries.)
This choice is somewhat arbitrarily chosen and based on the theory of the dynamo mechanism, which is one of the mechanisms thought to generate the powerful B-fields of magnetars \citep{tzw15}, as well as the requirement of birth spin periods of order milliseconds if magnetars are related to long $\gamma$-ray bursts, superluminous SNe and/or FRBs \citep[e.g.][]{ds21}.
For comparison, the investigations of \citet{bhvk19} use a constant value of $P_0 = 63\;{\rm ms}$ ($\Omega_0 = 100\;{\rm s}^{-1}$) while \citet{vrp+13} set $P_0=10\;{\rm ms}$. Nevertheless, as we shall see, our final results are not much dependent on the initial values of $P_0$.
For both the log-normal and the uniform probability distributions discussed above, we implement the SciPy statistics module \citep{SciPy}\footnote{\url{https://docs.scipy.org/doc/scipy/reference/stats.html}}.

\smallskip
\subsection{Fade-away}\label{Chapter3_FadeAway}
We take into account the emission properties of magnetars by introducing so-called fade-away. 
Using this procedure, we are able to determine which synthetic magnetars are detectable (visible) and which are non-detectable (faded). Only the visible ones are to be compared to the observed sample.
By fade-away, we refer to the process of a magnetar turning off. This happens when the mechanism driving the pulsed emission, be that X-ray or radio emission, ceases to function. 
In analogy, for radio pulsars fade-away is often modelled as a death line with a sharp cut-off in the $P\dot{P}$--diagram which separates the radio-loud NSs from the radio-quiet ones. The death line can be calculated from theory \citep{BeskinRadioPulsars,tbc+18} 
and is often included in investigations of radio pulsar evolution \citep{Ridley,gmvp14}.

Unfortunately, determining when magnetars fade is more complicated. This is due to the fact that the X-ray emission of magnetars is produced by losses of magnetic energy and is dependent on the poorly understood magnetospheric processes \citep{dvr+20}.
Finding an expression for fade-away would therefore require an in-depth analysis of the X-ray emission mechanisms, which is beyond the scope of this paper.
Instead, a simple analytic model, in which fade-away is a stochastic process, is employed.
Since the X-ray emission of magnetars is generated by losses of magnetic energy, the probability of fading, $P_{\rm fade}$, is chosen to be a function of the B-field strength \citep[however, see][]{dvr+20}.
We define:
\begin{equation}
\label{eq:P_fade}
    P_{\rm fade}(B) \equiv 1-S(B)\;.
\end{equation}
Here, $S(B)$ is the survival function, modelled as a log-logistic cumulative distribution function \citep[CDF, e.g.][]{FahimLogLog}:
\begin{equation}
     S(B) = \left( 1+\left(\frac{B}{s_1}\right)^{-s_2} \right)^{-1}\;,
\end{equation}
where the parameter $s_1$ is the median, while $s_2$ determines the steepness of the function. The initial values of these parameters are found by fitting $S(B)$ to the empirical CDF (ECDF) of the B-fields from the observed sample ($S(B)_{\rm obs}$) listed in Table~\ref{table:ObsMag}. 

\begin{figure}
\centering
\includegraphics[width=\linewidth]{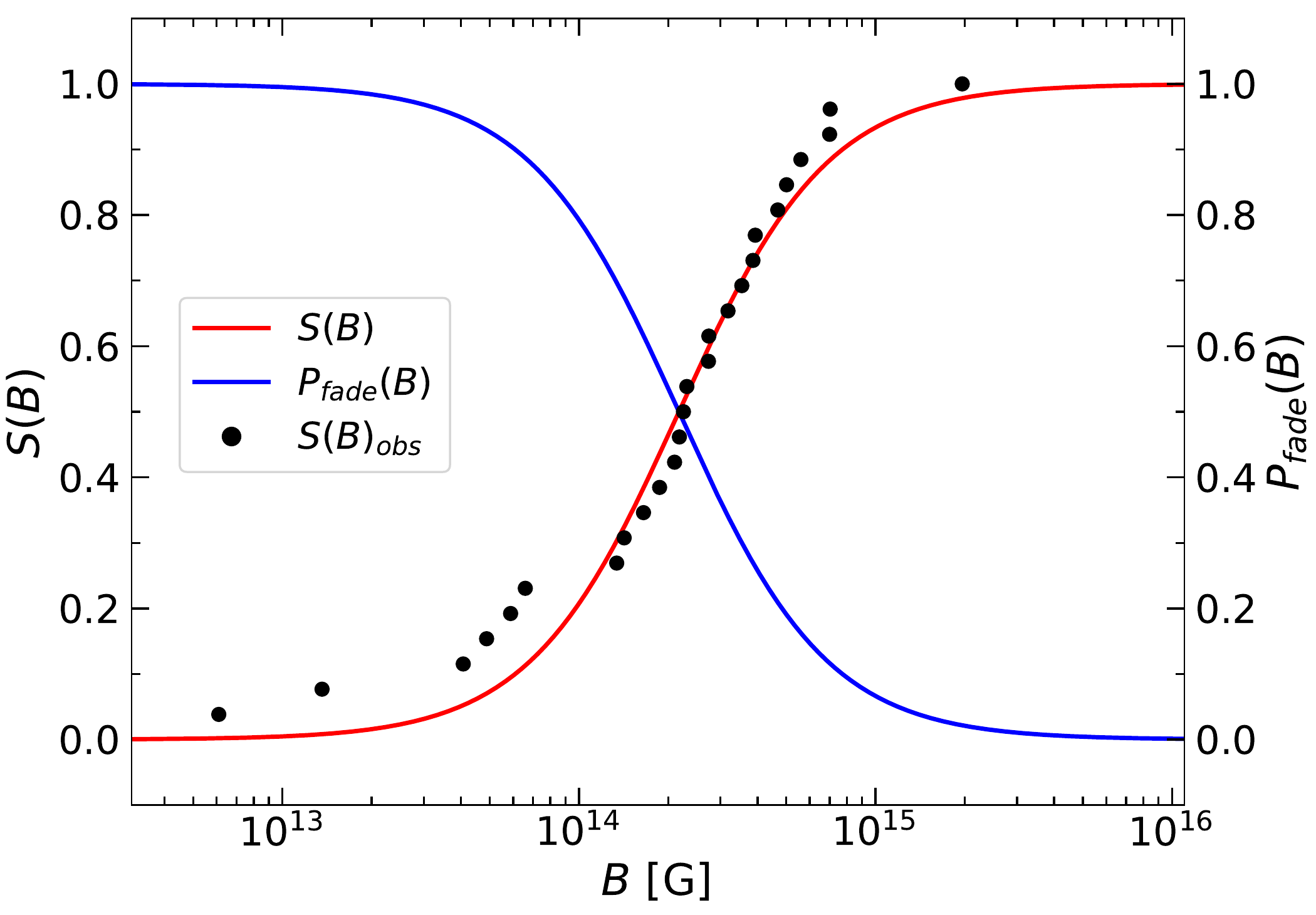}
\caption{The ECDF of the B-fields of the observed magnetars and the best fitting log-logistic CDF are plotted as black dots and a red line, respectively. The probability of fade-away, which is one minus the log-logistic CDF, is also plotted (blue).
Magnetar evolution is from right to left.}
\label{fig:loglogplot}
\end{figure}

Figure~\ref{fig:loglogplot} shows our fit, $S(B)$ to the data, $S(B)_{\rm obs}$ and the corresponding $P_{\rm fade}$. 
The values of $s_1$ and $s_2$ from the best fit are:
\begin{equation}
\label{eq:optimal_s1_s2}
s_1 = 2.17 \times 10^{14}\;{\rm G} \\ s_2 = 1.73
\end{equation}
It can be seen that the survival function follows $S(B)_{\rm obs}$ quite well, except at smaller values of $B$, where $S(B)_{\rm obs}$ lies above the log-logistic fit.
In general, we find that $s_2 = 1.73$ is usually a good choice while $s_1$ has to be varied based on the choice of evolutionary avenue and the decay timescale of the B-field.

Beaming is another phenomenon that influences which magnetars are visible. It is not taken into account here, we simply assume to first order a beaming factor of 1, i.e. all active magnetars are visible \citep[however see discussions in][]{oze01,oze02}.

\smallskip
\subsubsection{Dependence of fade-away on B-fields and inclination}\label{Chapter3_a_and_B}
The electromagnetic detectability of any astrophysical source depends on its emission properties over time. 
Fade-away, and thus the exact position and shape of the visible population in the $P\dot{P}$ diagram, is dependent on the way in which the B-fields of the synthetic magnetars are calculated. 
We investigate how the choice of the function used to calculate the crustal B-field, $B(t)$ and the distribution of initial magnetic inclination angles, $\alpha_0$ affect the non-faded synthetic magnetars.

We test two commonly used expressions for $B(t)$, namely:
\begin{align}
&B(t) = \begin{cases}
	B_{\rm dip} = \displaystyle\sqrt{\frac{P\dot{P}}{K \sin^2\alpha}},\\
	B_{\rm mag} = \displaystyle\sqrt{\frac{c^3 I P\dot{P}}{4 \pi^2 R^6}\frac{1}{1+\sin^2\alpha}}.
\end{cases}
\label{eq:B(t)}
\end{align}
The first equation ($B_{\rm dip}$) is simply the dipole-estimated B-field derived from equation~\ref{eq:Pdot}.
The second equation ($B_{\rm mag}$) is derived from a model which combines the vacuum dipole with a magnetosphere \citep{spi06}, see \citet{tlk12}. 
Due to the inclusion of magnetoshperic effects, the dependence of $B_{\rm mag}$ on $\alpha$ is reduced, and a braking torque is present even if the spin- and the B-field axis of the NS are completely aligned.
We also test two different distributions of the initial magnetic inclination angle, $\alpha_0$: the uniform (i.e. flat probability), $\Phi_{\rm uni}(\alpha_0)$ and the sinusoidal distribution, $\Phi_{\rm sin}(\alpha_0)$ --- see Appendix~\ref{AppendixA} for details. 

We find that the choices of $B(t)$ and $\Phi(\alpha_0)$ do not matter much. Regardless of their combination, a visible synthetic population that matches observations can be achieved as long as other parameters are adjusted accordingly.
In the rest of this investigation, we choose to use $B(t)=B_{\rm dip}$ and $\Phi (\alpha_0)=\Phi_{\rm sin}(\alpha_0)$.

\begin{figure*}
\centering
\includegraphics[width=\linewidth]{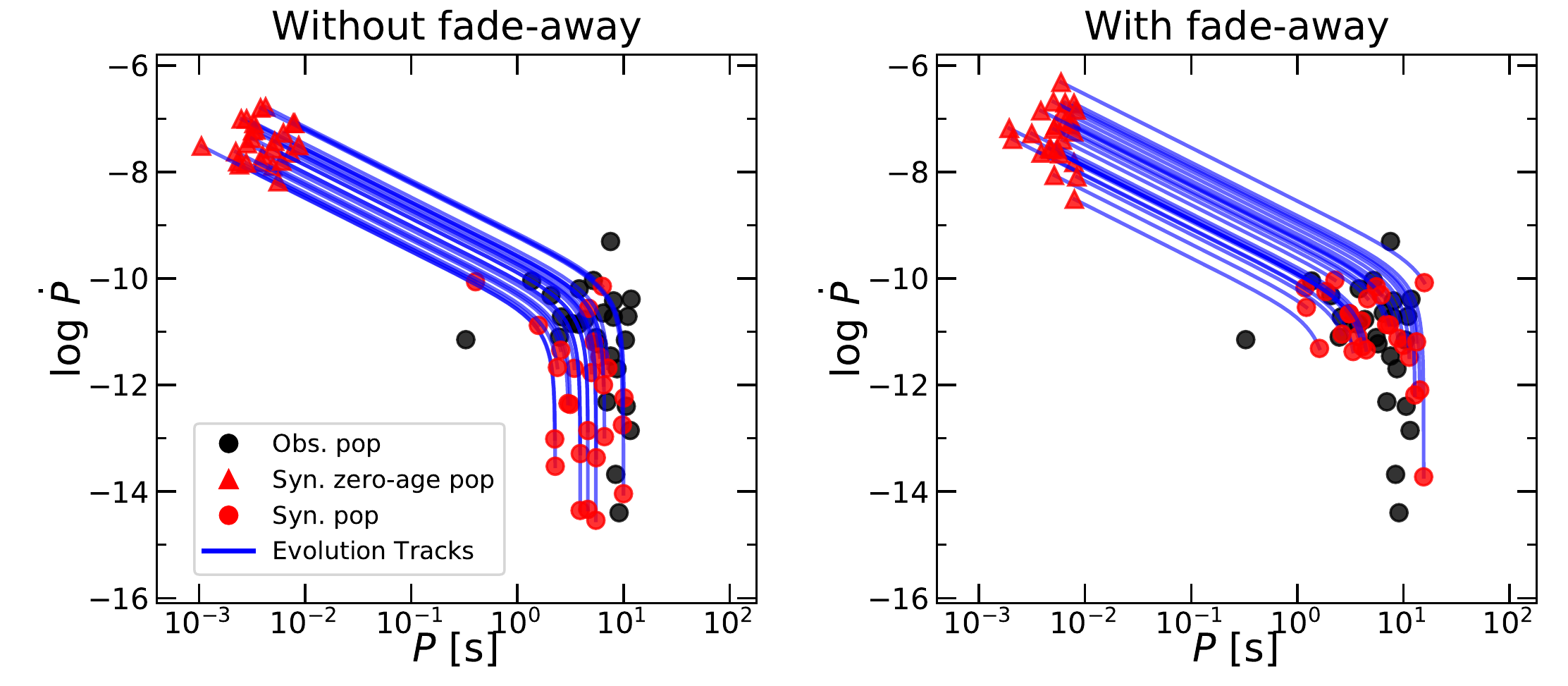}
\caption{Synthetic magnetar populations produced without (left) and with fade-away (right), based on iterations from the optimized model $A^M_3$ (Table~\ref{tab:manualmodels}), except here they are evolved with a constant B-field axis inclination of $\alpha=\alpha_0=\pi/2$. The outlier J1846$-$0258 (see also Fig.~\ref{fig:ObsMag}) is discussed in Section~\ref{subsubsec:zero-age-param}.}
\label{fig:Two_pops}
\end{figure*}

\subsection{Evolving a population}
To evolve our zero-age population, we assign a randomly-drawn true age between 0 and $t_{\rm max}$ to all our generated magnetars. 
Producing a final synthetic population of magnetars thus requires a full set of parameters specifying initial zero-age variables and those related to the evolutionary avenue. We therefore introduce the vector, $\vec{\theta}$, containing all of the required values:
\begin{align}\label{eq:theta-function}
&\vec{\theta} = 
\begin{cases}
	( \mu_{P_0}, \,\sigma_{P_0}, \,\mu_{\dot{P}_0}, \,\sigma_{\dot{P}_0}, \,t_{\rm max}, \,\tau_B, \,s_1)       & {\rm Avenue~A}\\
	( \mu_{P_0}, \,\sigma_{P_0}, \,\mu_{\dot{P}_0}, \,\sigma_{\dot{P}_0}, \,t_{\rm max}, \,\tau_{B}, \,\beta, \,s_1)   & {\rm Avenue~B}\;,
\end{cases}
\end{align}
After choosing an avenue and setting all parameter values in $\vec{\theta}$, a synthetic population is generated and evolved in the following way:
\begin{enumerate}
    \item Create a synthetic magnetar by generating $P_0$, $\dt{P}_0$, $\alpha_0$ and $t$ from the distributions in Table~\ref{tab:zero_age}.
    \item Evolve $P$ and $\dt{P}$ until true age $t$, using equations~(\ref{eq:Pdot}) and ~(\ref{eq:Both_avenues_P}), corresponding to the chosen avenue.
    \item Calculate the B-field at age $t$ using $B_{\rm dip}$ from equation~(\ref{eq:B(t)}).
    \item Find $P_{\rm fade}$ and determine if the magnetar is visible or faded.
\end{enumerate}
This cycle is repeated until the desired number of visible magnetars is generated. In the end, the synthetic population (containing $N_{\rm tot}$ magnetars) consists of two sub-populations: the visible population with $N_{\rm vis}$ magnetars, and the faded population consisting of $N_{\rm fad}=N_{\rm tot}-N_{\rm vis}$ magnetars. Only the former is detectable and thus it is the one that is compared to the observed magnetars.

\begin{figure}
\centering
\includegraphics[width=\linewidth]{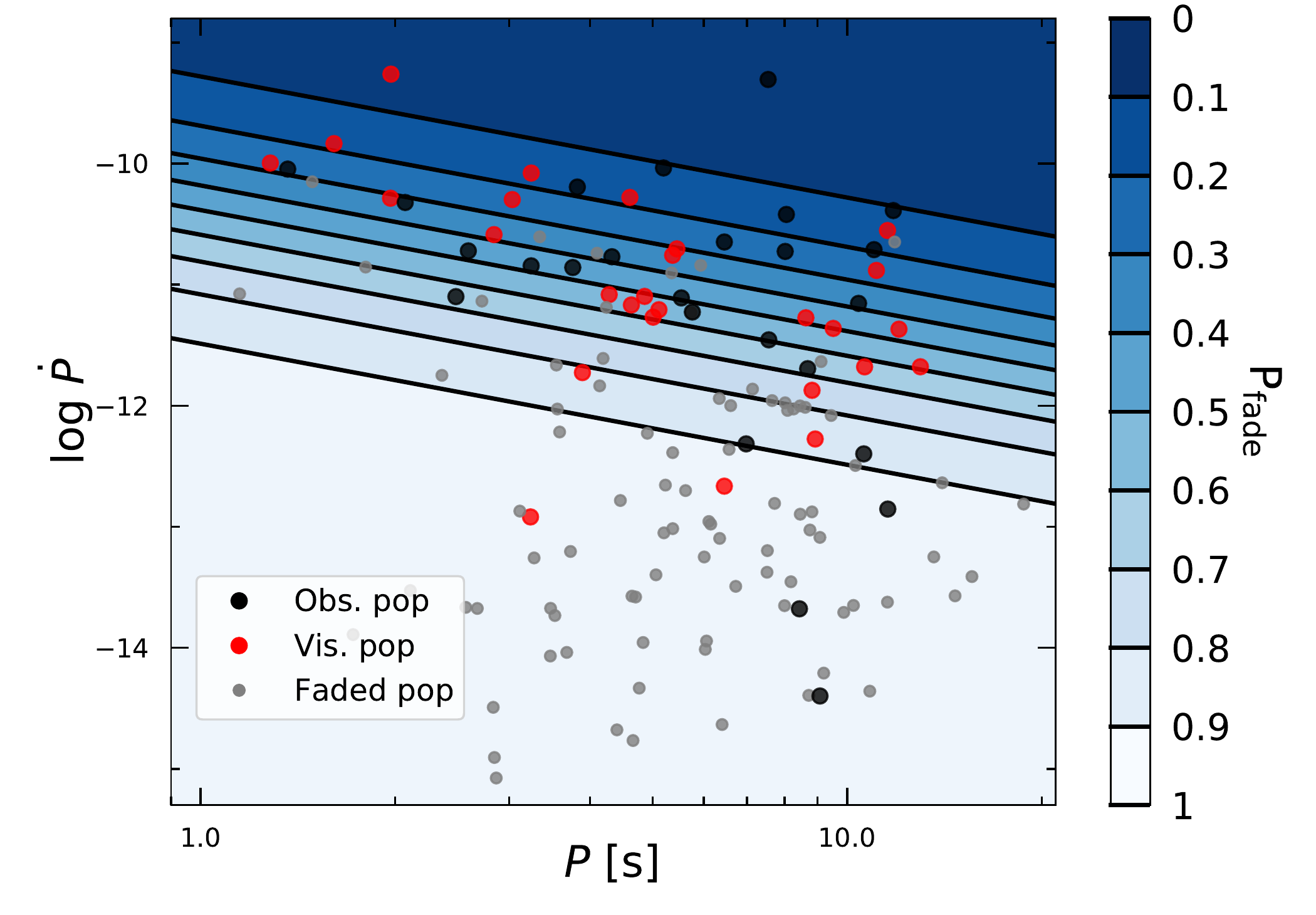}
\caption{Fade-away diagram for the magnetar population shown in the right panel of Fig.~\ref{fig:Two_pops}. For comparison: the visible, faded, and observed populations are plotted in: red, light grey, and black points, respectively. In addition, $P_{\rm fade}$ is plotted as a blue colour gradient. The black lines are probability contours $P_{\rm fade} = 0.1,\,0.2,...\, ,0.9$, corresponding to lines of constant B-field ($P_{\rm fade} = 0.1$ (top) and 0.9 (bottom) correspond to $B\simeq 1.3\times 10^{15}\;{\rm G}$ and $B\simeq 1.0\times 10^{14}\;{\rm G}$, respectively; see equation~\ref{eq:P_fade}).}
\label{fig:Two_pops_heatmap}
\end{figure}

Figure~\ref{fig:Two_pops} displays two populations of synthetic magnetars. The population in the right panel is an iteration of the optimized model~$A^M_3$ (see Table~\ref{tab:manualmodels} and Section~\ref{sec:optimizing}), with the exception that the initial magnetic angle, $\alpha_0$, is kept constant at $\alpha_0 = \pi/2$ for all synthetic magnetars. 
The population in the left panel is synthesized from the same model, but is evolved without fade-away. By keeping $\alpha_0$ constant, we can plot the probability of fade away in a $P \dt{P}$--diagram. This is shown in Fig.~\ref{fig:Two_pops_heatmap}, where $P_{\rm{fade}}$ is plotted as a color gradient together with the population from the right panel of Fig.~\ref{fig:Two_pops}. The black contours mark $P_{\rm fade}=\lbrace0.1,\,0.2,\,0.3,\,0.4,\,0.5,\,0.6,\,0.7,\,0.8,\,0.9\rbrace$.

Older magnetars with larger $P$ and smaller $\dot{P}$ (i.e. smaller surface B-fields) end up in the area of the $P\dot{P}$--diagram where $P_{\rm fade}$ is larger.
This causes a larger fraction of them to fade and disappear from the observable population.
In this way, the visible population ends up consisting mostly of magnetars with larger $\dot{P}$ values, which causes it to resemble the observed population much better than the synthetic population evolved without fade-away. 

Varying the $s_1$ and $s_2$ parameters changes the shape and position of $P_{\rm fade}$ in $P\dot{P}$-space. Increasing (decreasing) $s_2$ leads to $P_{\rm fade}$ becoming steeper (less steep), causing the contours of equal $P_{\rm fade}$ to converge (diverge). Choosing $s_2 \gg 1$ yields a hard-limit death line.
On the other hand, increasing (decreasing) $s_1$ moves $P_{\rm fade}$ up (down) along the lines of constant $B$.

Our results with fade-away show a general tendency to somewhat underproduce magnetars in the tail region. The reason for this discrepancy can be understood from the deviation between the log-logistic CDF and the observed magnetars with lower B-fields plotted in Fig.~\ref{fig:loglogplot}. This issue affects all our models and cannot be fixed by simply varying $s_1$ and $s_2$. Choosing a different fitting function, $S(B)_{\rm obs}$ could possibly help alleviating this problem.

Lastly, another effect of including fade-away to the evolution procedure is that $t_{\rm max}$ stops playing a large role since above a certain true age, magnetars are almost certain to fade. 
Despite this, $t_{\rm max}$ has to be kept as a free parameter. It has to be varied alongside $\tau_{B}$ so that the synthetic magnetars have the chance to evolve past the bend (``knee'') in their evolutionary track (see e.g. Fig.~\ref{fig:Evo_tracks_both_avenues}). If $t_{\rm max}<\tau_{B}$, then the decay timescale plays no role and the evolution is essentially that of a constant B-field (constant $n$). Although, such a scenario is physical, it cannot reproduce the observed population of magnetars, see model $A_0^M$ in Fig.~\ref{fig:Full_plots_Appendix_1}.

\section{Optimizing Models}\label{sec:optimizing}
Previously when discussing synthetic populations, we stated that some populations ``fit well to the observed one'', purely based on qualitative visual inspection in the $P\dot{P}$--diagram. 
In the following, we apply the Kolmogorov-Smirnov (K-S) test as a mean to determine the goodness of fit. The results of such tests are then used to optimize synthetic populations.

Due to limited computational resources, we optimize models where some of the parameters contained in $\vec{\theta}$ (equation~\ref{eq:theta-function}) are kept constant. Optimization thus refers to varying the remaining free parameters until we find the $\vec{\theta}$ that on average yields the best fitting populations of visible magnetars ($\vec{\theta}_{\rm{opt}}$). 
A variety of our models are shown in Tables~\ref{tab:manualmodels} and \ref{tab:automodels}. 
The model nomenclature relates to the applied evolutionary avenues.

Besides evaluating the fit to the observed magnetar population in the $P\dot{P}$--diagram, another important parameter is the resulting birth rate (BR) of magnetars. A hard upper limit on the BR is $20\;{\rm kyr}^{-1}$ \citep{dhk+06}, which is achieved assuming conservatively that {\em all} Galactic core-collapse supernovae (CCSNe) produce magnetars, i.e. here disregarding the possibility of CCSNe producing pulsars that are only radio-loud or black holes. Thus all our models that end up with ${\rm BR}>20\;{\rm kyr}^{-1}$ are considered unrealistic \citep[in analogy with][]{kk08,bhvk19}.

We use two different algorithms to optimize models: manual, and automatic algorithms. Both utilize the K-S test and we start by introducing this test method.

\subsection{Kolmogorov-Smirnov (K-S) test}\label{subsec:KS1D}
The Kolmogorov-Smirnov (K-S) test is a simple statistical test applicable to empirical measures of independent variables. It utilizes the largest difference between the ECDFs of the compared measures as the test statistic. See Appendix~\ref{AppendixD} for a detailed description of the two-sample K-S test.

Two separate K-S tests are performed: one for the distribution of $P$ and the other for $\dt{P}$. The p-values (between 0 and 1) from these tests serve as an indicator of goodness of fit, higher values signifying better correspondence between the observed and visible populations.
Following \citet{Ridley}, we define a figure of merit, FOM, to serve as an overall indicator:
\begin{equation}
  \label{eq:FOM_definition}
  \text{FOM} = (1-\text{p-value}(P)) + (1-\text{p-value}(\dot{P}))
\end{equation}
and thus $0\le {\rm FOM} \le 2$. The $\vec{\theta}$ which on average yields the lowest FOM is considered to be the optimal choice for a particular model.

\begin{figure}
\centering
\makebox[\columnwidth][c]{\includegraphics[width=1.1\linewidth]{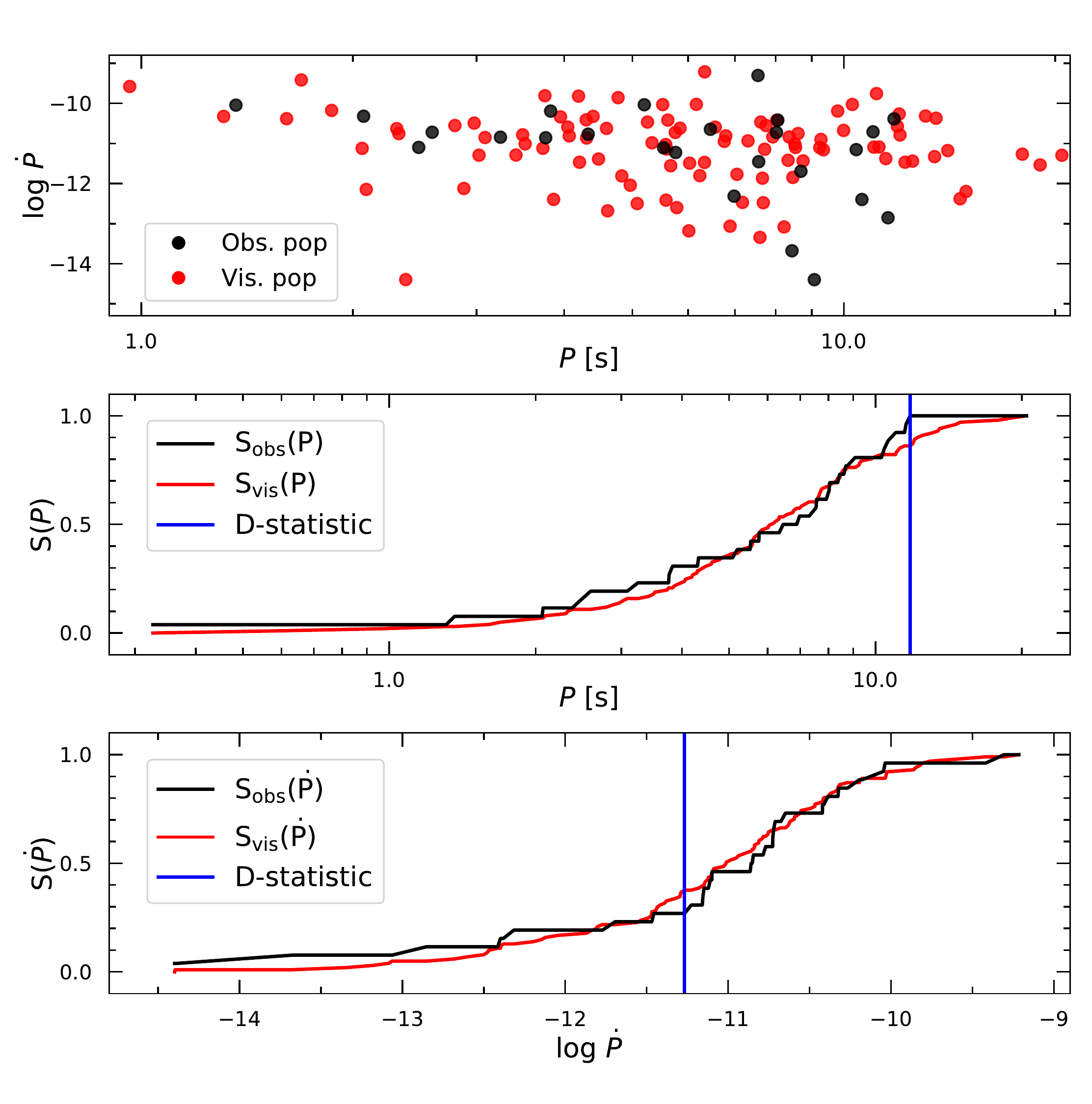}}
\caption{Top panel: $P\dot{P}$--diagram of 100 synthesized visible magnetars (red dots) from an iteration of model $A^M_3$ with $\tau_B = 5\;{\rm kyr}$. Only the visible magnetars are plotted to reduce clutter. 
Central panel: $S_{\rm vis}(P)$ and $S_{\rm obs}(P)$.
Bottom panel: $S_{\rm vis}(\dot{P})$ and $S_{\rm obs}(\dot{P})$.  
The positions of largest differences between the ECDFs are marked with blue vertical lines.}
\label{fig:KS_example1}
\end{figure}

Figure~\ref{fig:KS_example1} is a plot of the $P\dot{P}$--diagram containing a synthetic visible magnetar population. This population is an iteration of the optimized model $A^M_3$ defined in Table~\ref{tab:manualmodels},  containing 100 visible magnetars.
The ECDFs of $S_{\rm vis}(P)$, $S_{\rm obs}(P)$, $S_{\rm vis}(\dot{P})$ and $S_{\rm obs}(\dot{P})$ 
are plotted in the central and bottom panels of the figure. The locations of largest differences between the ECDFs are marked with blue vertical lines. Our resulting K-S tests yield: $\text{p-value}(P)=0.765$, $\text{p-value}(\dot{P})=0.947$, and thus $\text{FOM}=0.288$.

\smallskip
\subsection{Manual optimization algorithm}\label{subsubsec:manual}
Having defined the FOM, we can proceed to model optimization. After a model is defined and all constants are set, we search for the optimal $\vec{\theta}$ by manually varying free parameters following an algorithm based on the one used by \citet{fk06}. This algorithm is described in Appendix~\ref{AppendixC}.

\begin{table*}
\begin{tabular}{ccrrrcccc}
\toprule
Model & $\vec{\theta}$ & $t_{\rm max}$ & $\tau_B$  & $\beta$ &  $\mu_{\dt{P}_0}$ & $s_1$  & $\mean{\rm FOM}$ & $\mean{\rm BR}\;$\\
      &                & (kyr)         & (kyr)     & (kyr)        &  & (G) & & (${\rm kyr}^{-1}$)\\
\midrule
$A_0^M$ & $\{ \mu_{\dt{P}_0},\,s_1 \}$ & 20 & $\infty$   & --- & $9.000\times 10^{-9}$  & $1.000\times 10^{15}$    &$1.574{\pm 0.013}$ & $7.985{\pm 0.044}$\\
$A_1^M$ & $\{ \mu_{\dt{P}_0},\,s_1 \}$ &  5 & 0.5        & --- & $4.525\times 10^{-7}$  & $8.814\times 10^{13}$    &$1.023{\pm 0.016}$ & $16.192{\pm 0.088}$\\
$A_2^M$ & $\{ \mu_{\dt{P}_0},\,s_1 \}$ &  8 & 1          & --- & $2.356\times 10^{-7}$  & $9.915\times 10^{13}$    &$0.821{\pm 0.017}$ & $9.345{\pm 0.046}$\\
$A_3^M$ & $\{ \mu_{\dt{P}_0},\,s_1 \}$ & 25 & 5          & --- & $4.051\times 10^{-8}$  & $3.610\times 10^{14}$    &$0.653{\pm 0.016}$ & $6.169{\pm 0.036}$\\
$A_4^M$ & $\{ \mu_{\dt{P}_0},\,s_1 \}$ & 50 & 10         & --- & $1.842\times 10^{-8}$  & $8.737\times 10^{14}$    &$0.747{\pm 0.017}$ & $11.395{\pm 0.070}$\\
$A_5^M$ & $\{\mu_{\dt{P}_0},\,s_1 \}$ &  2 & 0.2 & --- & $1.000\times 10^{-6}$  & $6.000\times 10^{13}$   & $1.303 {\pm 0.017}$ & $32.951 {\pm 0.161}$\\
$A_6^M$ & $\{\mu_{\dt{P}_0},\,s_1 \}$ & 100 & 20  & --- & $9.000\times 10^{-9}$  & $4.000\times 10^{15}$   & $1.159 {\pm 0.016}$ & $79.43 {\pm 0.52}$\\
\\
$B_0^M$ & $\{ \mu_{\dt{P}_0},\,s_1 \}$ & 10      & 5  & $-0.5$ & $6.695\times 10^{-8}$  & $9.085\times 10^{13}$  &$0.729{\pm 0.019}$ & $4.152{\pm 0.016}$\\
$B_1^M$ & $\{ \mu_{\dt{P}_0},\,s_1 \}$ &  5       & 5  & $-1$ & $7.119\times 10^{-8}$  & $2.154\times 10^{12}$  &$0.895{\pm 0.021}$ & $5.212{\pm 0.002}$\\
$B_2^M$ & $\{\mu_{\dt{P}_0},\,s_1 \}$ & 100  & 5   & 0.5 & $3.000\times 10^{-8}$  & $3.000\times 10^{15}$   & $0.735 {\pm 0.019}$ & $57.07 {\pm 0.35}$\\
\bottomrule
\end{tabular}
\caption{Optimal parameter values and results for 10 models optimized using the manual algorithm. Parameters contained in $\vec{\theta}$ are kept free and varied during optimization. All models use: $\mu_{P_0} = 0.005\;{\rm s}$, $\sigma_{P_0}=0.5$,  $\sigma_{\dot{P}_0}=1$, and $s_2=1.73$. Note that the $\mean{\rm FOM}$ and $\mean{\rm BR}$ values are averages from 1000 iterations with  $\vec{\theta}_{\rm opt}$. The lower the value of $\mean{\rm FOM}$, the better the fit (see Section~\ref{subsec:KS1D} and Appendix~\ref{AppendixD} for details).}
\label{tab:manualmodels}
\end{table*}

Table~\ref{tab:manualmodels} lists the results of this optimization procedure. They are marked with a superscript $M$ (manual) in order to distinguish them from the models optimized using the automatic algorithm (superscript $A$ in Table~\ref{tab:automodels}). 
Due to the stochastic nature of zero-age populations, the FOM for a specific choice of parameters can vary quite a lot --- see Fig.~\ref{fig:FOMhist} where a histogram of 1000 FOMs is plotted. 
Therefore, many population runs are required for each parameter choice to find the average FOM ($\mean{\rm{FOM}}$) and the average BR ($\mean{\rm{\rm{BR}}}$). The listed values of the $\mean{\rm{FOM}}$ and $\mean{\rm{BR}}$ are results from runs performed using $\vec{\theta}_{\rm opt}$,  $N_{\rm vis} = 100$ and $N_{\rm ite} = 1000$ (Appendix~\ref{AppendixC}).
Table~\ref{tab:manualmodels} also includes models resulting in a poor fits due to either large $\mean{\rm{FOM}}$ and/or too large $\mean{\rm{BR}}$. These are models: $A_0^M$, $A_5^M$, $A_6^M$, and $B_2^M$. 

\smallskip
\subsection{Automatic optimization}\label{subsubsec:automatic}
Manual optimization yields quite good synthetic populations. However, it is very inefficient in exploring the space of free parameters, and there is no certainty that the found $\vec{\theta}_{\rm opt}$ represents a true global minimum in $\mean{\rm{FOM}}$. Producing a wide grid of parameters is out of the question, as the bad parameter choices yield populations with many faded magnetars per visible one. This can take a long time to compute, even if a check is put into place that terminates models which exceed the $\mean{\rm{BR}}$ limit of $20\;{\rm kyr}^{-1}$. There is a need for a method that can efficiently sort out the bad parameter choices and converge to regions with better solutions.
For this reason, we turn to the automatic optimization algorithm. 

This new algorithm is based on the one used by \citet{gmvp14}, which itself is built upon the two-dimensional annealing method described in \citet{PressKS}. However, unlike \citet{gmvp14}, we apply the one-dimensional K-S test and minimize the $\mean{{\rm FOM}}$. The algorithm is explained in detail in Appendix~\ref{AppendixC}.
Unlike the manual method, this approach is somewhat automated as it uses a random walk process to explore the space of free parameters. With every cycle, the limits of the space of free parameters are narrowed down until a minimum in the $\mean{{\rm FOM}}$ is reached.

\subsubsection{Automatic models}\label{subsec:_automodels}
Table~\ref{tab:automodels} lists our two selected automatic models ($A^A$ and $B^{A}$) and their results. 
These models are started with very wide limits on $\vec{\theta}$, chosen such that they could reproduce all of the models listed in Table~\ref{tab:manualmodels} that have a $\mean{\rm{BR}}$ below $20~\rm{kyr}^{-1}$.

Figure~\ref{fig:2D_results} shows visible magnetar populations from the optimized automatic models.
Furthermore, in Appendix~\ref{AppendixC}, we show the $\mean{{\rm FOM}}$ plots from the first and the last optimization cycle of model $A^A$.

\begin{table*}
\begin{tabular}{ccrrrcccc}
\toprule
Model & $\vec{\theta}$ & $t_{\rm max}$ & $\tau_B$ & $\beta$ &  $\mu_{\dt{P}_0}$ & $s_1$  & $\mean{\rm FOM}$ & $\mean{\rm BR}\;$\\
      &                & (kyr)         & (kyr)     & (kyr)        &  & (G) & & (${\rm kyr}^{-1}$)\\
\midrule
$A^A$ & $\{ \mu_{\dt{P}_0},\,s_1,\,\tau_B \}$ & 50 & $3.952$ & ---  & $5.366\times 10^{-8}$  & $2.871\times 10^{14}$    &$0.634{\pm 0.016}$ & $5.897{\pm 0.036}$\\

$B^A$ & $\{ \mu_{\dt{P}_0},\,s_1,\,\tau_{B},\,\beta \}$ &  50 &  4.155   & -0.2581  & $6.662\times 10^{-8}$  & $1.298\times 10^{14}$    & $0.664{\pm}0.017$ & $4.470{\pm 0.022}$\\
\bottomrule
\end{tabular}
\caption{Optimal parameter values and results for two models optimized using the automatic algorithm. Both models use: $\mu_{P_0} = 0.005\;{\rm s}$, $\sigma_{P_0}=0.5$, $\sigma_{\dot{P}_0}=1$, and $s_2=1.73$.
Note that the $\mean{\rm FOM}$ and $\mean{\rm BR}$ values are averages from 1000 iterations with $\vec{\theta}_{\rm opt}$.}
\label{tab:automodels}
\end{table*}

\begin{figure*}
\centering
\makebox[\textwidth][c]{\includegraphics[width=\linewidth,height=0.333\textheight]{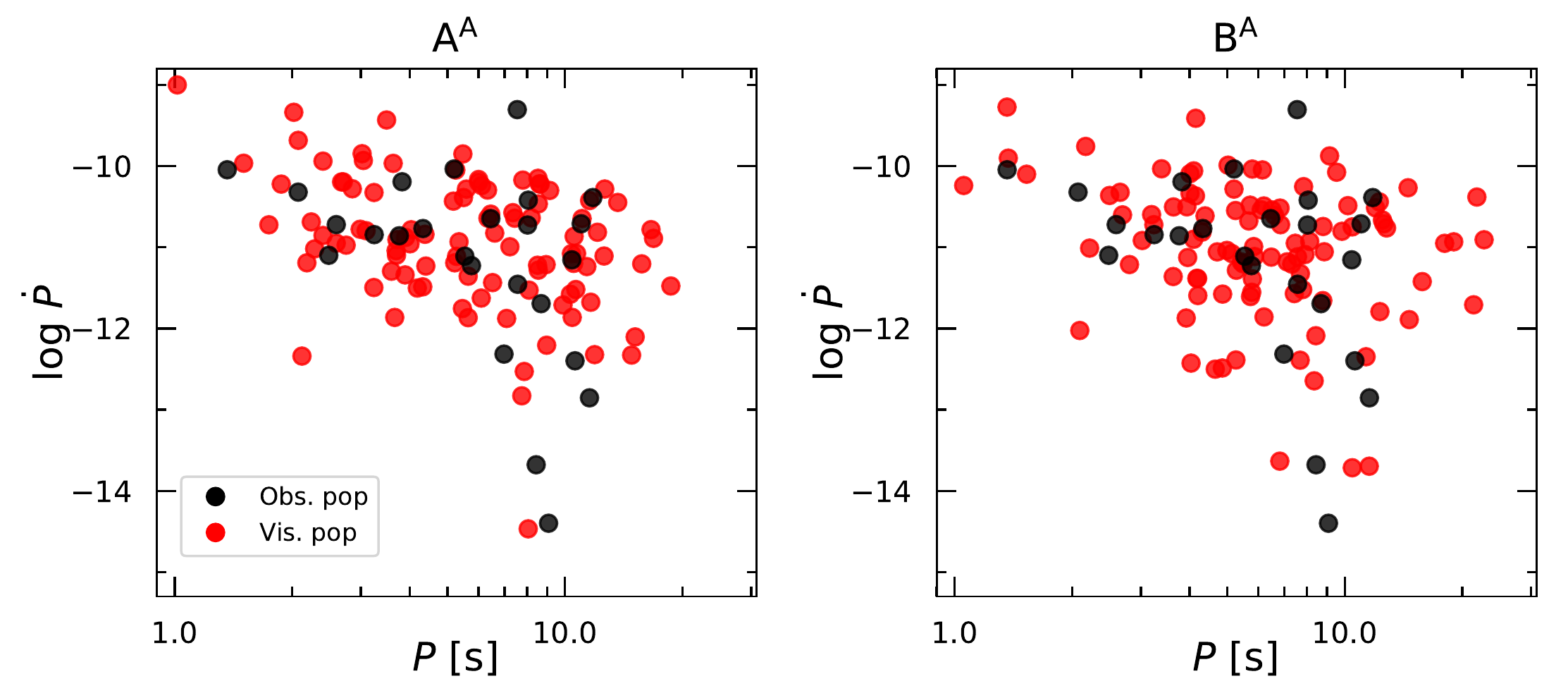}}
\caption{$P\dot{P}$--diagrams of synthetic populations that are optimized iterations of the two models listed in Table~\ref{tab:automodels}.}
\label{fig:2D_results}
\end{figure*}\textbf{}

\section{Discussion}\label{sec:discussion}
The following discussion is split into three sections. We start by discussing the merits and drawbacks of the two optimization algorithms. 
Then, we consider the values of parameters that can account for the observed population and yield a $\mean{\rm{BR}}$ less than the upper limit of $20\;{\rm kyr}^{-1}$. Note that this means that models $A_5^M$, $A_6^M$ and $B_3^M$ are not taken into account in this section.
Finally, the effects of fade-away are evaluated before moving on to a brief general discussion of future work and improvements. Our summary follows thereafter.

\subsection{Optimization algorithms}
The two algorithms used to optimize models yield quite similar results. Unfortunately, both of them have drawbacks.
On one hand, the manual algorithm is poor at exploring the space of free parameters. This proves especially to be a problem when optimizing B-models, as the additional degree of freedom in $\beta$ makes it harder to conclude if a true minimum in $\mean{{\rm FOM}}$ is reached.

On the other hand, when using the automatic algorithm it quickly becomes hard to identify a clear minimum in $\mean{{\rm FOM}}$. Already after 4--5 cycles the minimum becomes significantly less pronounced. This effect became worse with further cycles until we ended up with wide ranges of free parameters yielding rather similar values of $\mean{{\rm FOM}}$ (see Appendix~\ref{AppendixC}). Due to this effect, it is difficult to determine whether a true global minimum has been achieved.
We also tried limiting the size of $\vec{\theta}$ and start with solutions 
close to $\vec{\theta}_{\rm opt}$ from the manual models. This results in the automatic algorithm converging faster. However, if initiated in such a way, the algorithm usually ended up converging to some $\vec{\theta}_{\rm opt}$ not too far away from the manual solution.  This family of models is therefore not included in the paper. 

To conclude, we find that the application of automatic algorithm was not necessary for producing well fitting synthetic populations. The $\mean{{\rm FOM}}$s for the solutions found by the automatic optimization are lower than any of the manual models, but due to the limitations of $\mean{{\rm FOM}}$ as an indicator of the goodness of fit, we do not consider this to be a significant improvement.

\subsubsection{Finding the ``best'' model and effect of stochasticity}
Our initial presumption was that there existed some set of free parameters which would produce the best synthetic populations, ones that most closely resemble the observed population in the $P\dot{P}$--diagram.

As stated in Section~\ref{sec:optimizing}, reviewing $P\dot{P}$-diagrams of the synthetic populations is not sufficient to evaluate goodness of fit, as synthetic populations are quite random. 
Although introducing the FOM as a measure of the goodness of fit of a specific choice of parameters and producing multiple populations  helps to alleviate this issue, the FOM is not perfect. 
Since the zero-age populations are generated from probability distributions and the fade-away procedure is stochastic, synthetic populations and their FOM vary a lot, even if they are generated using the same $\vec{\theta}$ (see Fig.~\ref{fig:FOMhist}). It is for this reason that we performed 1000 iterations of each specific set of model parameters.
Furthermore, as described in Appendix~\ref{AppendixD}, the K-S test is limited by the small sample size of observed magnetars and is also hindered by the lacking sensitivity to outliers \citep{PressKS,gmvp14}. Thus, the FOM does not change much if the synthetic and observed population differ in the border regions, such as the tail.

This, combined with the aforementioned drawbacks of the manual and automatic algorithms, makes it difficult to settle on a single ``best'' model.
Still, the K-S test is quite good at determining when the fit is poor. For this reason, we switched our goal to determining how different parameters affect the synthetic populations and finding the limits beyond which it becomes impossible to reproduce the observed magnetars, either due to too large $\mean{\rm{BR}}$ or $\mean{\rm{FOM}}$. 

A more precise measure of the goodness of fit would improve the analysis in future work.
Employing a different statistical test might therefore possibly significantly improve the usefulness of the FOM as a measure of the goodness of fit. One possibility is a use a modified K-S test which is more sensitive to outliers \citep[e.g.][]{MasonModKS,PressKS}.

\subsection{Constraining Parameters}

\begin{figure*}
\centering
\makebox[\textwidth]{\includegraphics[width=0.5\textwidth]{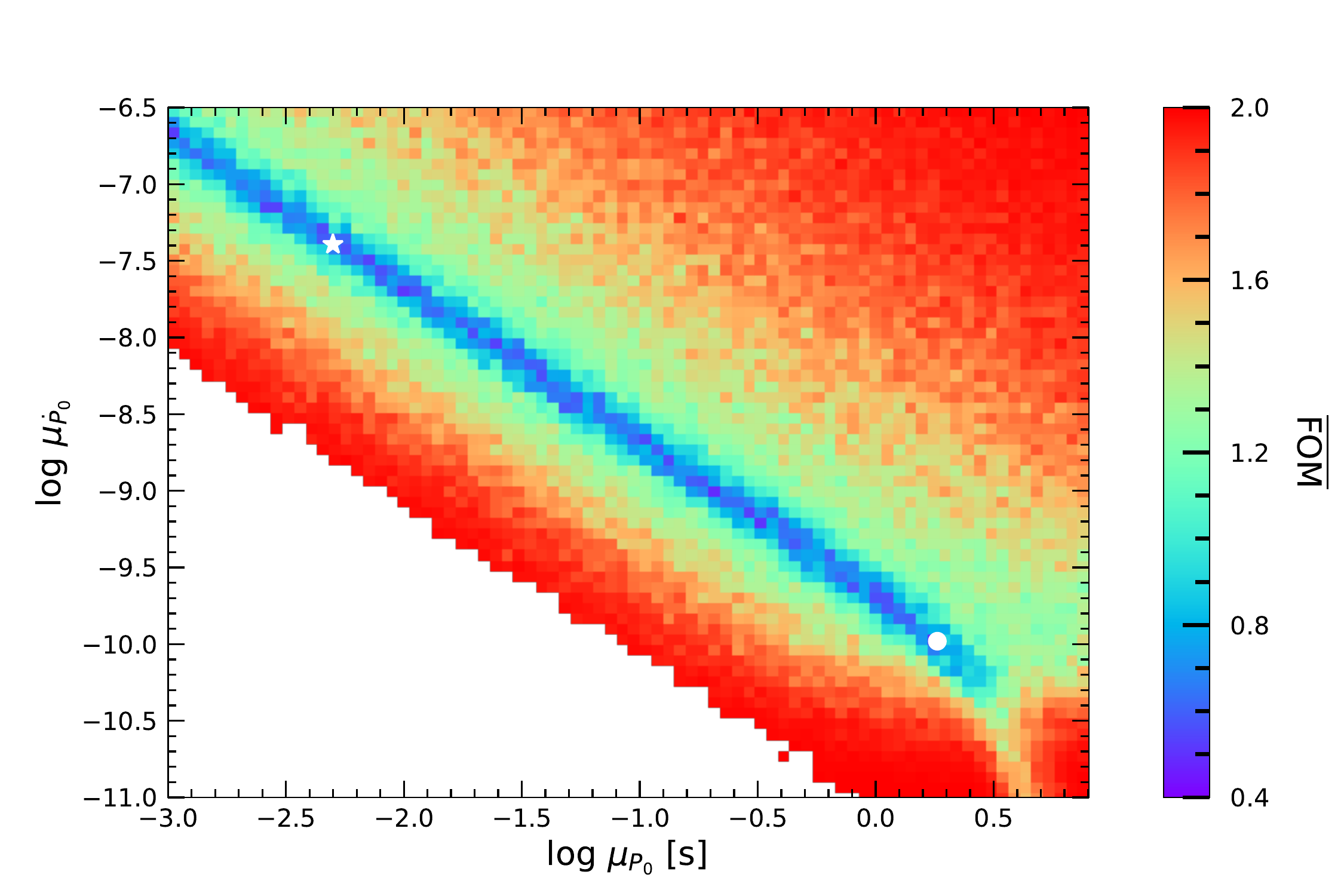}
\hspace{0.3cm} 
\includegraphics[width=0.5\linewidth]{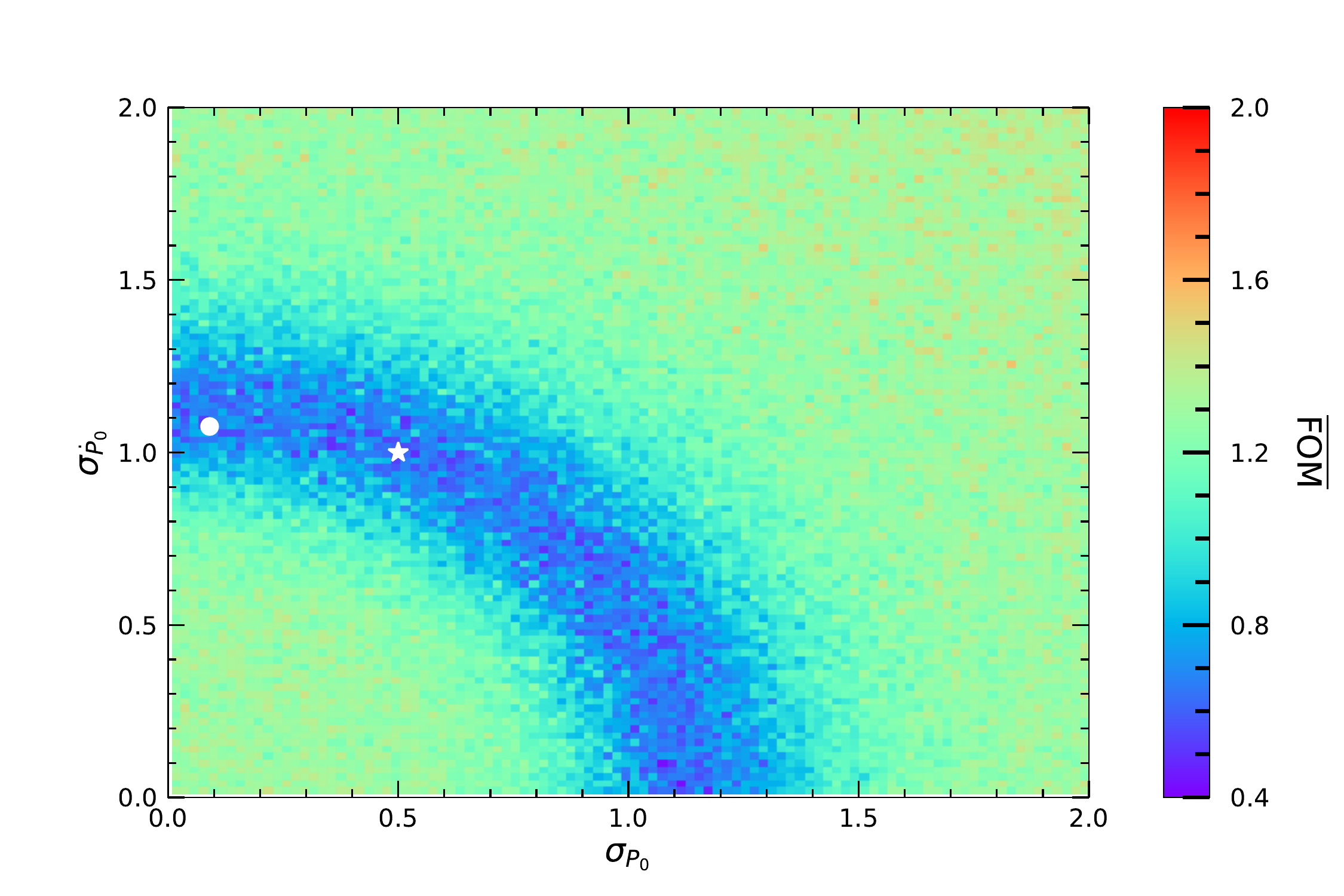}}
\makebox[\textwidth]{
\includegraphics[width=0.5\linewidth]{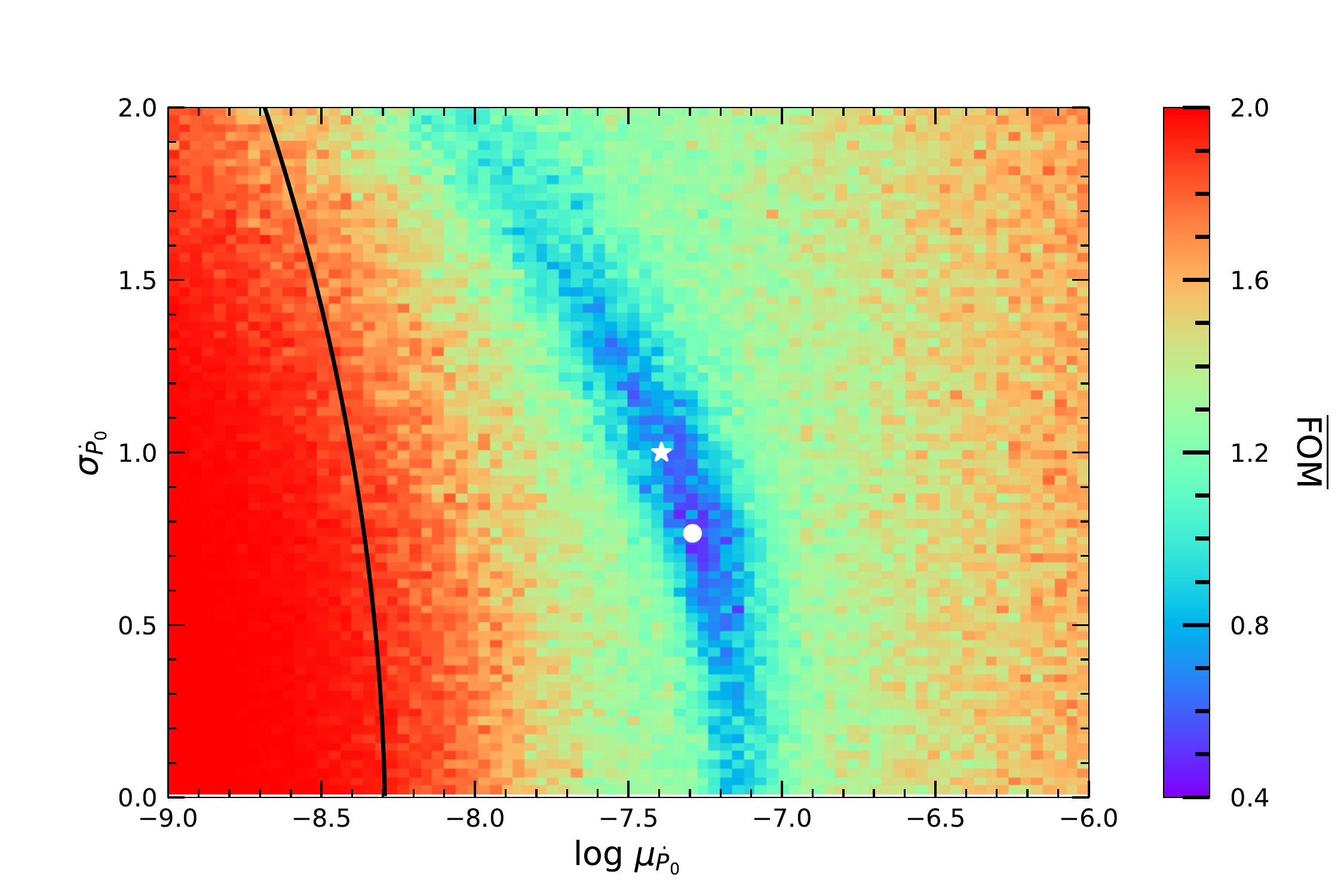}
\hspace{0.3cm}
\includegraphics[width=0.5\linewidth]{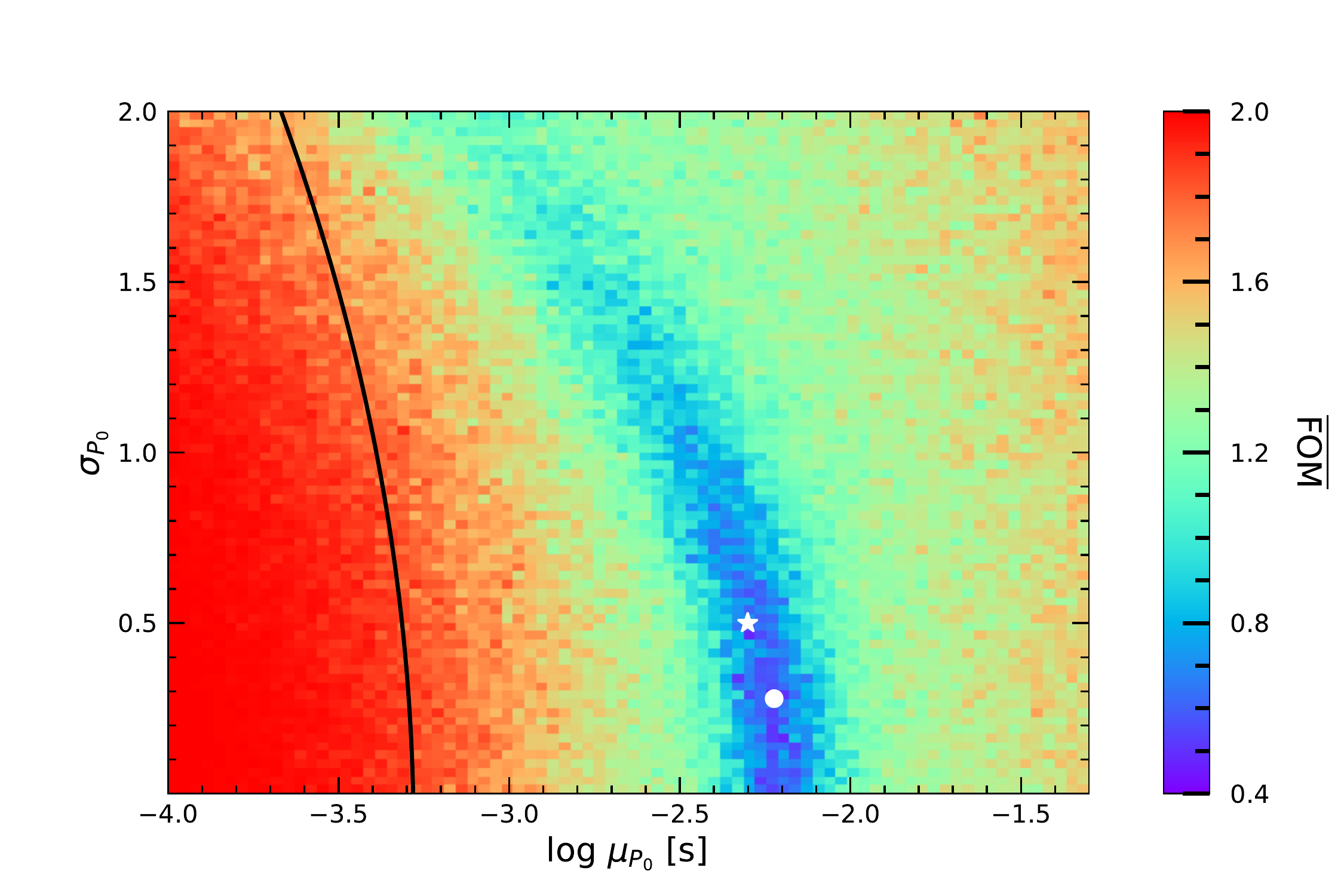}}
\caption{Heat maps of $\mean{{\rm FOM}}$ based on $\vec{\theta}_{\rm opt}$ of model $A_3$. For each panel, only the two parameters on the axes are varied. The white dots mark the solution with the lowest $\mean{{\rm FOM}}$ while the stars are the reference value for model $A^M_3$ listed in Table~\ref{tab:manualmodels}.
Upper left: $\log\mu_{P_0}$--$\log\mu_{\dot{P}_0}$ plane. Solutions in the white area were not calculated due to limited computational resources. They are all poor fits with $\mean{\rm{FOM}} = 2$. 
Upper right: $\sigma_{P_0}$--$\sigma_{\dot{P}_0}$ plane.
Lower left: $\log\mu_{\dot{P}_0}$--$\sigma_{\dot{P}_0}$ plane.
Lower right: $\log\mu_{P_0}$--$\sigma_{P_0}$ plane.
Solutions left of the black lines have $\mean{\rm{BR}}>20\;{\rm kyr}^{-1}$.}
\label{fig:muP_mudP_Cplot}
\end{figure*}

The results of the optimization procedures are used to constrain the space of parameters which can successfully account for the observed population of magnetars. We start by taking a look at the zero-age parameters used to generate newborn magnetars.

\subsubsection{Zero-age parameters}\label{subsubsec:zero-age-param}
In general, synthetic zero-age populations can be shaped differently and still end up reproducing the observed magnetars. To illustrate how changes in $\mu_{P_0}, \mu_{\dot{P}_0}, \sigma_{P_0}$ and $\sigma_{\dot{P}_0}$ affect the goodness of fit, we plot in Fig.~\ref{fig:muP_mudP_Cplot} four heat maps of $\mean{{\rm FOM}}$ around $\vec{\theta}_{\rm opt}$ of model $A^M_3$. The reference value for model $A_3^M$ is marked with a white star, while the white dot is the solution for the parameters that yield the smallest $\mean{{\rm FOM}}$ for the shown grid.

Starting with the upper left panel of Fig.~\ref{fig:muP_mudP_Cplot}, where a line of good solutions in $\mu_{P_0}$-$\mu_{\dt{P}_0}$ space is plotted. We find that, as long as $\mu_{P_0}<2\;{\rm s}$, there are many choices of $\mu_{P_0}$ and $\mu_{\dt{P}_0}$ which yield good synthetic populations.
For this reason, limiting $P_0$ to $1.6-16\;{\rm ms}$/ (Section~\ref{subsec:ZeroAge}) is not an issue.
This conclusion is in agreement with many other works on pulsar evolution \citep{fk06,gmvp14,bhvk19}.
Specifically, \citet{gmvp14}, who focus on radio pulsar evolution, find that $P_{0}=0.5\;{\rm s}$ is the upper limit. This is lower than our limit, due to the radio pulsar population possessing smaller spin periods.

In the upper right panel of Fig.~\ref{fig:muP_mudP_Cplot}, we plot $\sigma_{\dt{P}_0}$ versus $\sigma_{P_0}$. As long as the sum of these two parameters is in the range $1.0-1.5$, the resulting fit is equally good. 
However, the choice of $\sigma_{P_0}=0.5$ and $\sigma_{\dt{P}_0}=1$ that we made, has two main consequences.
Firstly, it becomes hard to replicate the outlier PSR~J1846$-$0258. Increasing the $\sigma$ parameters can account for the existence of this outlier, but it also leads to an overall larger $\mean{\rm{FOM}}$ (i.e. worse fit). In particular, when fade-away is introduced, models with $\sigma$ parameters large enough to reliably reproduce PSR~J1846$-$0258 often require BRs exceeding $20\;{\rm kyr}^{-1}$.
We find it tempting to suggest that PSR~J1846$-$0258 is a result of a distinct evolutionary path, in agreement with the finding of \citet{bhvk19}. However, there are still effects that could explain the disparity between PSR~J1846$-$0258 and the rest of the magnetar population, which are not taken into account here. Some of these are: differences in mass,  non-dipolar B-field configuration, and envelope composition \citep{shs17,nk11}.

Secondly, replicating the tail of the observed population becomes hard. As shown in Figs.~\ref{fig:Full_plots_Appendix_1} and \ref{fig:Full_plots_Appendix_2} (in Appendix~\ref{AppendixB}), for the optimized models the synthetic populations tend to be too broadly distributed in the tail region. Choosing smaller $\sigma$ parameters could prevent this, but this results in a worse fit to the bulk of the observed population. I.e. the synthetic populations have an equal width in the bulk and tail regions, while the observed population narrows down in the tail. 
This is partially fixed by introducing fade-away. However, this combined behaviour cannot be replicated perfectly using the considered models. Either the models have to be modified to allow the synthetic magnetars to converge when they get older, or perhaps the observed sample is incomplete and is missing relatively many objects from the tail region. 

Finally, from the two bottom panels of Fig.~\ref{fig:muP_mudP_Cplot}, we find, similarly to~\citet{gmvp14}, that there are trends in the $\mu_{P_0}$--$\sigma_{P_0}$  and $\mu_{\dt{P}_0}$--$\sigma_{\dt{P}_0}$ planes. In both cases, there is a valley of decent solutions which bends towards smaller $\sigma$ as $\mu$ increases.
This effect can be explained in the following way.
When $\mu_{P_0}$ ($\mu_{\dt{P}_0}$) is shifted to smaller values, the centre of the synthetic population starts to move towards smaller $P$ ($\dot{P}$).
However, due to the dispersion in $\Phi(P_0)$ ($\Phi(\dot{P}_0)$), some magnetars are still produced in the region which evolves towards the observed population. 
The lower the $\mu_{P_0} $($\mu_{\dt{P}_0}$) parameter gets, the larger the $\sigma_{P_0}$ ($\sigma_{\dt{P}_0}$) parameter has to be in order to allow for the production of synthetic magnetars with large enough $P_0$ ($\dot{P}_0$) to fit to the observed population.
However, fade-away sorts the synthetic magnetars based on their estimated dipole B-fields, and it will therefore tend to eliminate the magnetars born with lower values of $P_0$ and $\dot{P}_0$ from the visible population (since $B\propto\sqrt{P\dot{P}}$). Thus, only the magnetars that evolve towards the observed population are left. This is why the valley of low $\mean{\rm{FOM}}$ solutions bends downwards with increasing $\mu_{P_0} $($\mu_{\dt{P}_0}$) in these plots.

The smaller $\mu_{P_0} $($\mu_{\dt{P}_0}$) becomes the larger the BR, as the visible magnetars essentially become outliers of the much larger total population. Eventually the BR will exceed the critical limit of $20\;{\rm kyr}^{-1}$. 
This limit is marked with a black line on the two bottom panels of Fig.~\ref{fig:muP_mudP_Cplot}.

Finally, once the value of $\mu_{P_0} $($\mu_{\dt{P}_0}$) reaches the region that evolves towards the observed population, it cannot be increased more without a loss in $\mean{\rm{FOM}}$. A larger $P_0$ ($\dot{P}_0$) would produce magnetars that do not fit the observed population and cannot be removed by fade-away, as they would have too large B-fields.

\subsubsection{Evolutionary avenues and parameters}
In the manually optimized models presented in this work, we wary the decay timescale of the B-field, $\tau _B$ across the Avenue~A-models (Table~\ref{tab:manualmodels}). For Avenue~B-models, we vary $\beta$ instead.
It is evident that the decay of the B-field is necessary since models with constant B-fields, like $A_0^M$, cannot be reconciled with the observed population of magnetars. Even with fade-away, they do not reproduce the tail and they tend to not fit the bulk well either (see Fig.~\ref{fig:Full_plots_Appendix_1}). This is no surprise, of course, since high-energy emission in X-rays and $\gamma$-rays from magnetars is believed to be powered by B-field decay.
For the models with finite decay time ($\tau_B \neq \infty$), the observed population can be accounted for as long as $0.5 \la \tau_B \la 10\;{\rm kyr}$. Setting $\tau_B<0.5\;{\rm kyr}$ or $\tau_B>10\;{\rm kyr}$ results in $\mean{{\rm BR}}$ increasing beyond the critical value of $20\;{\rm kyr}^{-1}$ (see Table~\ref{tab:manualmodels}). In addition, $\mean{{\rm FOM}}$ increases to such a degree that these models are certainly worse fits. This is also true for Avenue~B-models.

Interestingly, we find that the lowest $\mean{\rm{FOM}}$ (and $\mean{\rm{BR}}$) is found when $\tau_B \approx 4\;{\rm kyr}$. This is lower than the $\tau_B = 10\;{\rm kyr}$ found by \citet{bhvk19} and \citet{cgp00} as their best fit to empirical data. Notice however, our model~$A_4^M$ which also has $\tau_B = 10\;{\rm kyr}$ and lies within the acceptable BR limits.
In contrast, \citet{vrp+13} find $\tau_B$ to be at least $100\;{\rm kyr}$. In our analysis, models with such large values of $\tau_B$ end up with much too large BR~values.

Comparing Avenues~A and B, we can see that choosing $\beta >0$ is always a poor choice. This conclusion is in agreement with \citet{bhvk19} who found this choice of $\beta$ to yield very bottom-heavy populations. Due to the inclusion of fade-away in our analysis, it is $\mean{\rm BR}$ which becomes critically large. A synthetic population from one such model, $B_2^M$ with $\beta=0.5$, is plotted in Fig.~\ref{fig:Full_plots_Appendix_2}. This model results in $\mean{\rm BR}$ of $57.07{\pm 0.35}\;{\rm kyr}^{-1}$, much beyond the upper limit of $20\;{\rm kyr}^{-1}$ based on the Galactic CCSN rate. This discrepancy only gets worse the larger $\beta$ becomes.

On the other hand, choosing $\beta < 0$ is a viable choice. Such super-exponentially decaying B-fields result in synthetic populations that are top-heavy and have relatively fewer visible magnetars with weak B-fields. This outcome may also seem counter-intuitive.
The reason for the small amount of visible magnetars in the tail region is in this case the steep decrease of the B-field, see Fig.~\ref{fig:Params_both_avenues}. Since the decay happens super-exponentially, the window of true ages, which correspond to visible magnetars with weak B-fields, is very narrow (i.e. the evolution through the tail region of the $P\dot{P}$--diagram is very rapid). 

Among our super-exponentially decaying models, model $B_1^M$ is certainly the most remarkable one. It matches quite well with the observed population, even without fade-away. We did include fade-away in model $B_1^M$, but it can be seen in Fig.~\ref{fig:Full_plots_Appendix_2} that it was not strictly necessary. It is no surprise that \citet{bhvk19}, who do not consider a fade-away mechanism, find a model with $\beta=-1$ to best fit the observed population.

Avenue~B-models with $\beta <0$ are incomplete in nature since they cause the B-field to become negative. Thus, they cannot constitute a realistic description of spin evolution on their own.
A likely explanation is that equation~(\ref{eq:B}) only describes part of the NS B-field: i.e. in a model with a crustal and a core B-field, the former could be the one decaying while the latter is essentially constant on the considered timescale. In such a model, when the crustal field decays sufficiently, the long-term evolution would become dominated by the core component \citep{vrp+13}.

\smallskip
\subsubsection{Dependence on the NS equation-of-state (EoS)}\label{subsubsec:EoS}
For the constant introduced in equation~(\ref{eq:Pdot}), we chose as our default value $K=8.77\times 10^{-40}\;{\rm cm\,s^3\,g^{-1}}$ (e.g. corresponding to the case of a NS radius of $R=10\;{\rm km}$, and a moment of inertia, $I=1.11 \times 10^{45}\;{\rm g\,cm}^2$).   
If instead, we chose a significantly larger value of $K=11.2\times 10^{-40}\;{\rm cm\,s^3\,g^{-1}}$ (e.g. corresponding to a NS radius of $R=15\;{\rm km}$) and optimize in $(\mu_{\dt{P}_0},\;s_1)$ space, while keeping $\tau_B$ and other parameters at the values from the optimized model $A^A$ (Table~\ref{tab:automodels}), we find: 
$\mu_{\dt{P}_0}=6.020\times10^{-8}$, $s_1=1.698\times10^{14}$, $\mean{{\rm FOM}}=0.735\pm{0.018}$ and $\mean{\rm BR}=7.044{\pm 0.042}\;{\rm kyr}^{-1}$ for $R=15\;{\rm km}$. 
Hence, the net result of increasing the K-value by $\sim 30$\% (i.e. increasing the NS radius, mimicking a more stiff EoS) is that the synthesized population will have more magnetars with smaller B-fields, and thus more faded magnetars, which requires a higher BR to match the observations.

\subsection{Fade-away}
Fade-away (to our knowledge included in a magnetar spin investigation for the first time) proved to be an important addition for all models, except B-models with $\beta \leq -1$. Without fade-away, we could not produce synthetic populations that fit the observed one in the $P\dot{P}$--diagram.
The cost of this was additional free parameters. Adding degrees of freedom is a guaranteed way of making any model fit any data set, thus fade-away has to be carefully examined in order to evaluate its necessity as a part of magnetar evolution.

In essence, due to the inclusion of fade-away, the visible populations from different models are degenerate in the $P\dot{P}$--plane. 
However, as can be seen in Table~\ref{tab:manualmodels}, fade-away also causes different models to end up having very different $\mean{\rm{BR}}$.
On one hand, since model $B_1^M$ does not require fade-away and has a low value of $\mean{{\rm FOM}}$ (i.e. a good fit), fade-away could be seen as a needless addition. 
On the other hand, including fade-away yields $\mean{{\rm FOM}}$ values lower than that of $B_1^M$. 

To conclude, we find that including fade-away enables $\mean{\rm{BR}}$ to play an important role in determining $\tau _B$. 
It is difficult to determine whether the way in which fade-away is implemented in our models is reasonable. We believe that it is a productive inclusion, as it makes it possible to explore the scenario where the observed magnetars are a subset of a larger, partially faded population. However, the approach taken here is very crude, and the choice of the fitting function is suboptimal (see Fig.~\ref{fig:loglogplot}). A more advanced approach would require calculating the fading function based on the emission physics of magnetars, similar to the way that the death line for radio pulsars is treated \citep{cr93}.

\subsubsection{Birth rates}
Knowledge on the true magnetar BR is crucial for resolving the BR issue for the general population of all NSs \citep{kk08} and also for illuminating the possible evolutionary transition from one specie of NSs to another \citep{kas10}.
For our findings specifically, a precise knowledge of the BR would break the degeneracy between the different models, which would significantly constrain the space of parameters.

\citet{bhvk19} find the magnetar BR to be $2.3-20\;{\rm kyr}^{-1}$, while 
\citet{kk08} list a couple of estimates: the most reliable one being $6^{+9}_{-3}\;{\rm kyr}^{-1}$ derived only from the observations of persistent magnetars. 
However, they also note that the BR could be as high as $20\;{\rm kyr}^{-1}$ in the case that the B-field decays --- which is strongly supported by our work presented here. Based on our simulations, however, we cannot conclude which value of BR between $\sim 4-20\;{\rm kyr}^{-1}$ is more likely.

It is important to notice that the BRs of the optimized models are merely lower limits as magnetar beaming is not taken into account (i.e. we have assumed a beaming factor of 1, meaning that all active magnetars are visible). The inclusion of beaming would certainly narrow down the range of parameters that yield ${\rm BR} < 20\;{\rm kyr}^{-1}$. 
Finally, one has also to keep in mind that some SNe give birth to black holes, thus reducing the upper limit on the NS BR based on CCSNe.

\subsubsection{Faded populations}
If magnetars would produce other detectable NSs (such as RRATs or XDINSs; see Section~\ref{sec:intro}), measuring their $P$ and $\dt P$ would certainly be the best way to break the degeneracy between the models. We now discuss the possibility of faded magnetars being observable as RRATs or XDINSs.

\begin{figure*}
\centering
\includegraphics[width=\textwidth,height=0.444\textheight]{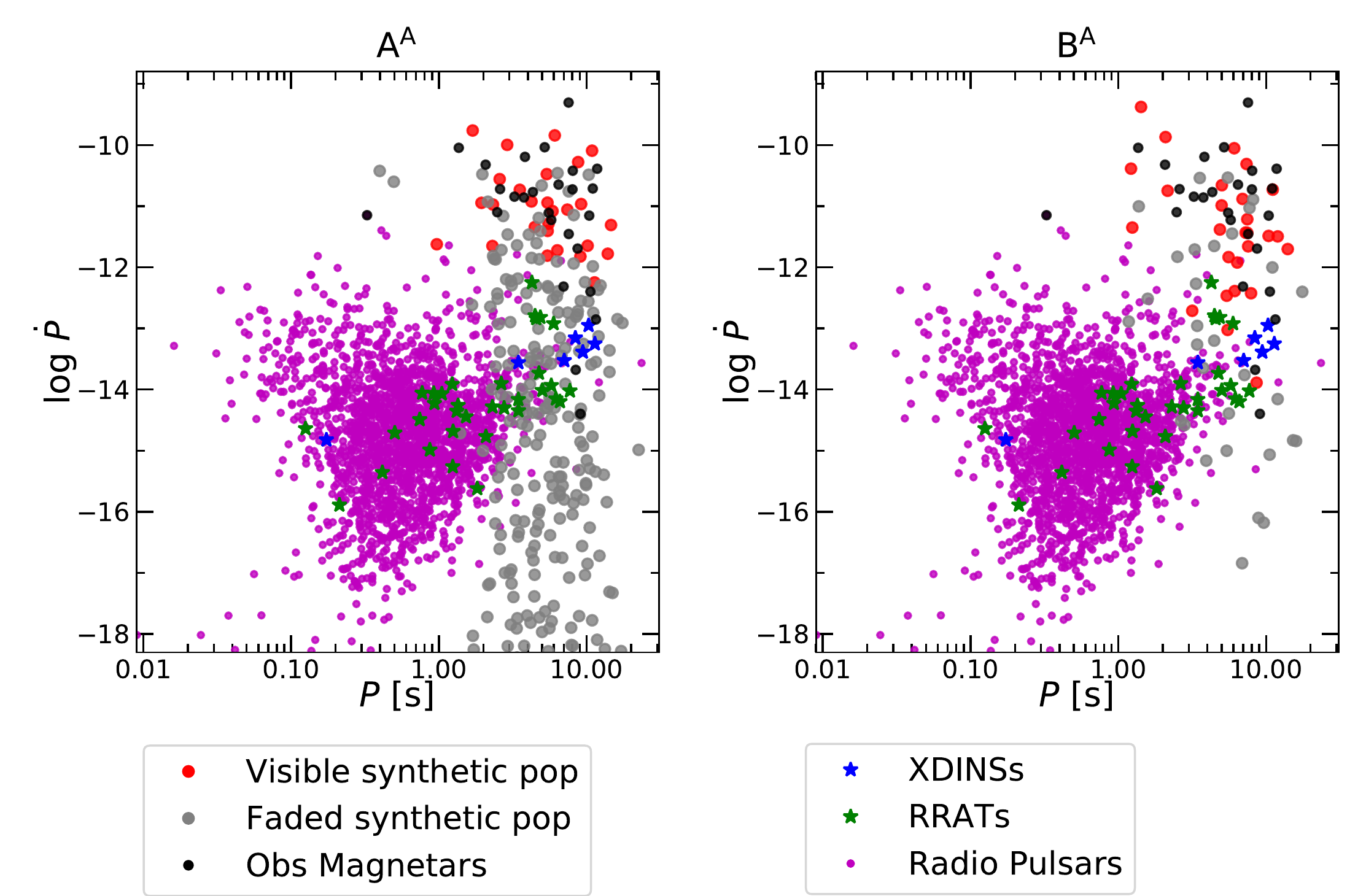}
\caption{$P\dot{P}$--diagram of visible and faded populations from the optimized models $A^A$ and $B^A$. RRATs, XDINSs and radio pulsars are also plotted --- see legend symbols. Data taken from the ATNF catalogue in May~2021: \url{http://www.atnf.csiro.au/research/pulsar/psrcat} \citep{mhth05}.}
\label{fig:P-dotP_wRRATS}
\end{figure*}

Figure~\ref{fig:P-dotP_wRRATS} shows a plot of the synthetic populations of magnetars from models $A^A$ and $B^A$ together with known radio pulsars, XDINSs and RRATs\footnote{The data is taken from the ATNF catalogue in May~2021.}.
The faded population of model $A^A$ is much more numerous than that of one of model $B^A$ due to the super-exponentialy decaying B-field of the latter. 
An issue for both models is that many faded magnetars end up having larger values of $\dt{P}$ compared to the XDINS and RRATS. 
More importantly, for both models, the RRATs cannot be fully reconciled with the faded population of magnetars, as many of them posses much lower spin periods. In this regard, XDINSs fit much better, i.e. begin located in the area containing the faded synthetic magnetars
\citep[see also][]{ppm+10} --- the only peculiar exception is the XDINS J1836+5925 which has $P=0.17\;{\rm s}$ \citep{aaa+09}.
Excluding model $A_0^M$, 6 out of the 7~XDINSs do exist in the region which is occupied by faded magnetars from our synthetic populations.
We conclude that an evolutionary link between magnetars and these other NS objects is possible, but this needs to be investigated further.
We notice that synthetic populations of faded magnetars vary quite a lot with $\beta$ and $\tau_B$ --- see Appendix~\ref{AppendixB} for further plots of faded populations. 

Finally, it should be mentioned that our evolutionary tracks of young NSs could, in principle, also connect to the location of the observed central compact objects (CCOs) in SN remnants (Fig.~\ref{fig:PdP_example}). As seen from Figs.~\ref{fig:Evo_tracks_both_avenues} and \ref{fig:Params_both_avenues}, this would require a very short decay timescale of a B-field which should not be too large at birth. However, these NSs would, by definition, not be magnetars. It has been hypothesized that CCOs are created from fallback of SN material \citep{gha13,mp95}.  

\section{Future work and summary}\label{sec:summary}
\subsection{Future work}\label{subsec:future}
To achieve further progress in understanding the spin (and B-field) evolution of the magnetar population, an improved comparison with data is needed. 
First of all, a better constrain on the the magnetar beaming factor and its possible dependence on $P$ and $B$ is needed \citep{oze01,oze02}. 
Secondly, using the K-S test is a good baseline, but it proves to be quite limited by producing somewhat uncertain FOMs.
A better statistical test, increased sample size of known magnetars, and a larger amount of trial and model iterations are important ingredients for future progress.

Another possibility is to use other magnetar observables to compare synthetic populations to the observed one.
The obvious choices are the X-ray luminosity, rates of bursts or giant flares, and magnetic energy losses \citep{gwkv00,kb17,bhvk19}. 
However, this requires reliable modelling of these phenomena and is thus difficult to implement for a trustworthy outcome.
One can also think of including knowledge on the Galactic location of magnetars, and thereby take into account kinematic properties and selection effects of magnetars --- similar to
work done on radio pulsars \citep{fk06,gmvp14}.

Moreover, one may consider further exploration of the decay of the magnetic inclination angle \citep{tm98,jk17} 
and/or take the magnetosphere \citep{spi06,ptl14} into account in the spin-down modelling. 
However, this would probably only change the torque by a factor of order unity, and thereby not reveal much new information on magnetar evolution in the $P\dot{P}$--diagram. 
Nevertheless, including the magnetosphere significantly changes the resulting distribution of magnetic inclination angles, $\alpha$. If only these angles could be constrained better from observations, one would have a tool to estimate the role of the magnetosphere. 

As mentioned already, the problem with our B-models with $\beta < 0$ is that the magnetar B-fields become negative at some point in their evolution. The simplest way to fix this issue is to add a constant core component ($B_{\rm core}$) to the decaying B-field:
\begin{equation}\label{eq:B_col_mod}
    B(t) = B_0 \left(1+ \frac{\beta t}{\tau_{B}} \right)^{\frac{-1}{\beta}} + B_{\rm core}.
\end{equation}
Such a model could be important for evaluating the possible evolutionary links between magnetars and other NSs, as it would significantly change the evolutionary tracks of old magnetars. 

Finally, considering magnetothermal evolution, \citet{vrp+13} predicts that the characteristic decay timescale is dependent on the initial B-field (e.g. as illustrated in their fig.~10). In all of the models considered here in our work, $\beta$ and $\tau_B$ are assumed to be identical for all magnetars.
Letting these two parameters be functions of the initial B-field would possibly allow for reproducing the results of \citet{vrp+13} and improve future investigations on the spin evolution of magnetars.
For example, it may be used to solve the problem of the large dispersion around the tails of the synthetic magnetar population in the $P\dot{P}$--diagram.

\subsection{Summary}
The aim of this work was to study the spin evolution of magnetars with a focus on determining the influence that their B-fields have on the evolution.
By considering the time dependence of the B-field decay, evolutionary tracks  (Fig.~\ref{fig:Evo_tracks_both_avenues}) were calculated and compared to observations. However, such tracks alone could not be used to determine how to best reproduce the observed magnetar population.
For this reason, synthetic populations were generated (Fig.~\ref{fig:Two_pops}) using two different evolutionary avenues (Avenue~A and Avenue~B), together with a novel fading procedure which accounted for the magnetars fading from detection as they age (Figs.~\ref{fig:loglogplot} and \ref{fig:Two_pops_heatmap}). 

By using two different optimization algorithms, a number of models were optimized (Tables~\ref{tab:manualmodels} and \ref{tab:automodels}), aiming to reproduce the observed population of magnetars and study the influence that the different parameters have on the synthetic populations.
Neither algorithm could find a single best model. The small sample size (26) of observed magnetars and the nature of the K-S test, such as the lack of sensitivity to outliers, limit the effectiveness of the FOM as an indicator of the goodness of fit.

Common for both A-models and B-models, we found that as long as the initial spin periods, $P_0$ are under $2\;{\rm s}$, the synthetic magnetars could be reconciled with the observed population. 
Thus, the $\mu$ and $\sigma$ parameters for the distributions of $P_0$ and $\dot{P}_0$ could be chosen in many different ways, as long as the upper limit was respected. This is consistent with the results of other works \citep{fk06,gmvp14,bhvk19}.
Due to fade-away, the $\mu$ and $\sigma$ parameters are correlated (Fig.~\ref{fig:muP_mudP_Cplot}, bottom panels). It was possible to produce models where the visible magnetars were outliers of a much larger total population. However, such models were disfavoured by the K-S test. 
The best results were obtained by setting $\sigma_{P_0}$ and $\sigma_{\dt{P}_0}$ as low as possible, with their sum being about $1.0-1.5$. 
We made the choice to use: $\mu_{P_0}=0.005\;{\rm s}$, $\sigma_{P_0}=0.5$, and  $\sigma_{\dt{P}_0}=1$ in all our final models. $\mu_{\dot{P}_0}$ and $t_{\rm max}$ had to be varied together with $\tau_B$.

In general, it was impossible to reproduce the peculiar outlier PSR~J1846$-$0258 (Fig.~\ref{fig:ObsMag}) without compromising the fit to the rest of the magnetars.
We thus come to the conclusion that this object is likely a product of a different evolutionary scenario, similar to the finding of \citet{bhvk19}.
If more similar objects are discovered, a serious reconsideration of formation paths and evolutionary models must be made. However, one should bear in mind that this source is also unique, being a magnetar initially detected as a radio pulsar.

The novel adoption of fade-away made it possible to account for the observed population in many different ways. Furthermore, it enabled us to disregard unrealistic models in which the BR is rising above $20\;{\rm kyr}^{-1}$, the total Galactic CCSN rate \citep{dhk+06}. In order to keep the BR below this value, $\tau_B$ had to be chosen between 0.5 and $10\;{\rm kyr}$. Although uncertain, the best results were obtained using $\tau_B$ close to $4\;{\rm kyr}$, which we therefore conclude is the typical decay timescale for the B-fields of magnetars, and thus their active lifetimes are similarly of order 4~kyr. 
This value is smaller by more than a factor of 2 compared to that obtained in previous works by \citet{cgp00,bhvk19}, who conclude that $\tau_B\simeq 10\;{\rm kyr}$.
In particular, we find that assuming the much longer decay timescales ($\sim 100-1000\;{\rm kyr}$) obtained from studies of magnetothermal evolution \citep{vrp+13}, we were not able to reproduce the observed population of magnetars.
It would be interesting to compare in more detail the direct evolutionary tracks from our model with those obtained from numerical magnetothermal models.

Choosing a sub-exponentially decaying B-field ($\beta > 0$) did not work, as the resulting population ended up having too large BRs.
On the other hand, super-exponential decay ($\beta < 0$) could reproduce the observed population. In general, we found that the viable values of $\beta$ range from $-1$ to $0$, although this range is sensitive to the chosen value of $\tau_B$.

The visible populations of synthetic magnetars from models in the aforementioned ranges of $\tau_B$ and $\beta$ ended up with very different $\mean{\rm{BR}}$ values. Therefore, a precise independent estimate of the $\mean{\rm{BR}}$ or a more thorough analysis of the faded populations could be used to narrow down the range of parameters. 

Comparing the faded synthetic population of magnetars to the XDINSs and RRATs was inconclusive (Fig.~\ref{fig:P-dotP_wRRATS}). Whereas most faded magnetars do end up overlapping with most XDINSs \citep{rip+13,vrp+13} and a number of RRATs, the faded magnetars often have much longer spin periods than the RRATS. 
They also extend to much smaller values of $\dot{P}$ than the observed XDINSs and RRATs do, although this may be an artefact of a too simple B-field decay models without considering the core B-field (Section~\ref{subsec:future}). 
Beware the relative number of faded magnetars (located across the Galaxy) and XDINS (located at $150-500\;{\rm pc}$) cannot be directly compared. 
Further analysis is required to confirm any evolutionary link between magnetars and other isolated NSs. 

Finally, we find that evaluating whether or not the sample of observed magnetars is complete is critical for the future success of the analysis \citep[see also][]{bhvk19}. The ranges of $\beta$ and $\tau_B$ values that account for the observed population could shift drastically if the fraction of missing magnetars is large and consists of objects with weak B-fields.

\section*{Acknowledgements}
We thank the referee for an insightful report that certainly improved our paper. JAJ and TMT acknowledge support from the Department of Physics and Astronomy (IFA) at Aarhus University.

\section*{Data availability}
The data underlying this article will be shared on reasonable request to the corresponding author.


\bibliographystyle{mnras}
\bibliography{magnetars}
\clearpage
\appendix

\section{Sample results from model $A_3$}\label{AppendixA}
\begin{figure*}
\centering
\includegraphics[width=0.90\textwidth]{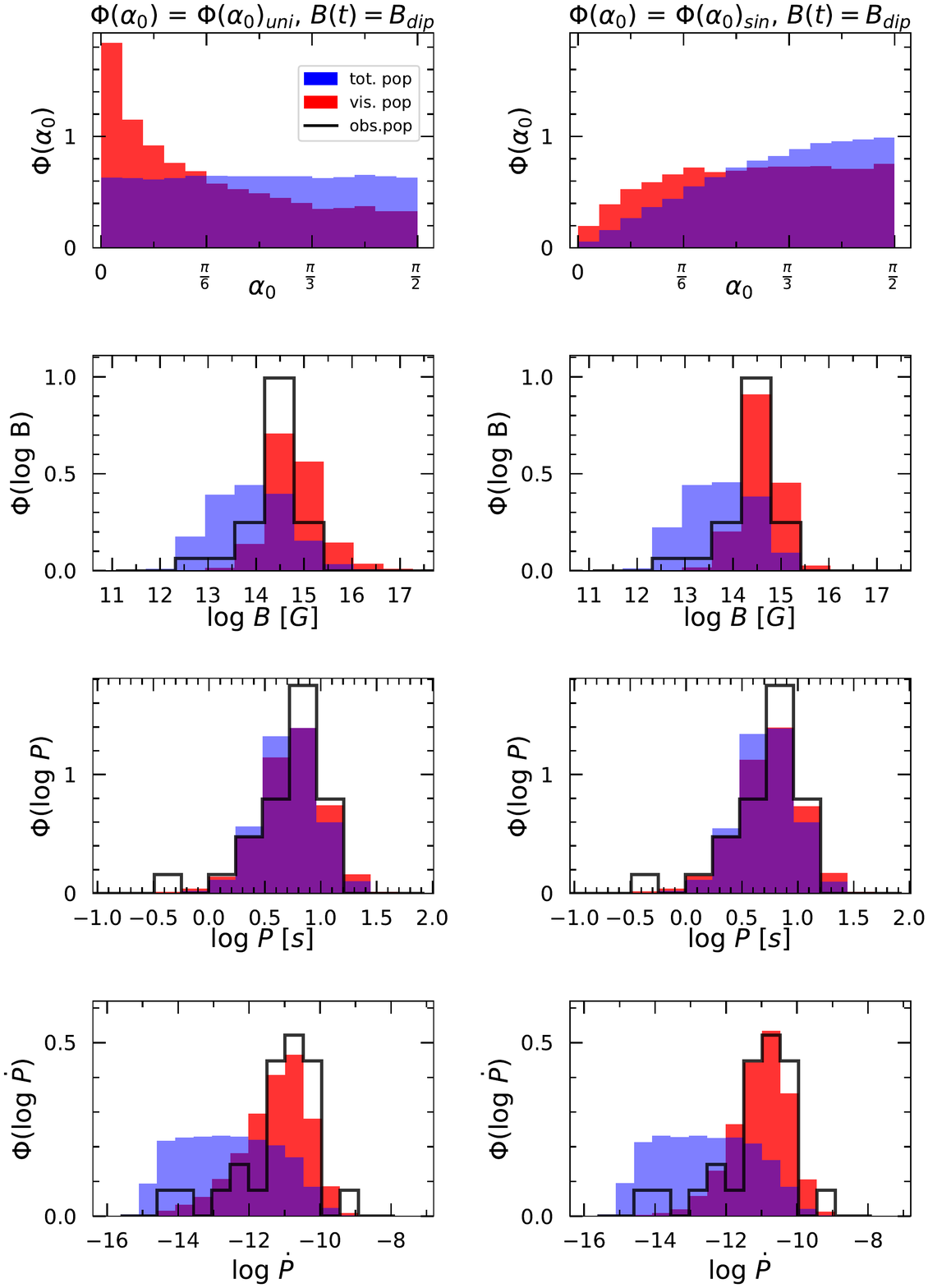}
\caption{Distributions of (top to bottom): magnetic inclination angles, $\Phi(\alpha_0)$; surface B-fields, $\Phi(B(t))$; spin periods, $\Phi(P)$; and spin period derivatives, $\Phi(\dot{P})$, for a synthetic population of magnetars for which $B(t)=B_{\rm dip}$. The left (right) column is for $\Phi (\alpha_0)$ being uniform (sinusoidal) --- see text.}
\label{fig:aB_hist1}
\end{figure*}

\begin{figure*}
\centering
\includegraphics[width=0.90\textwidth]{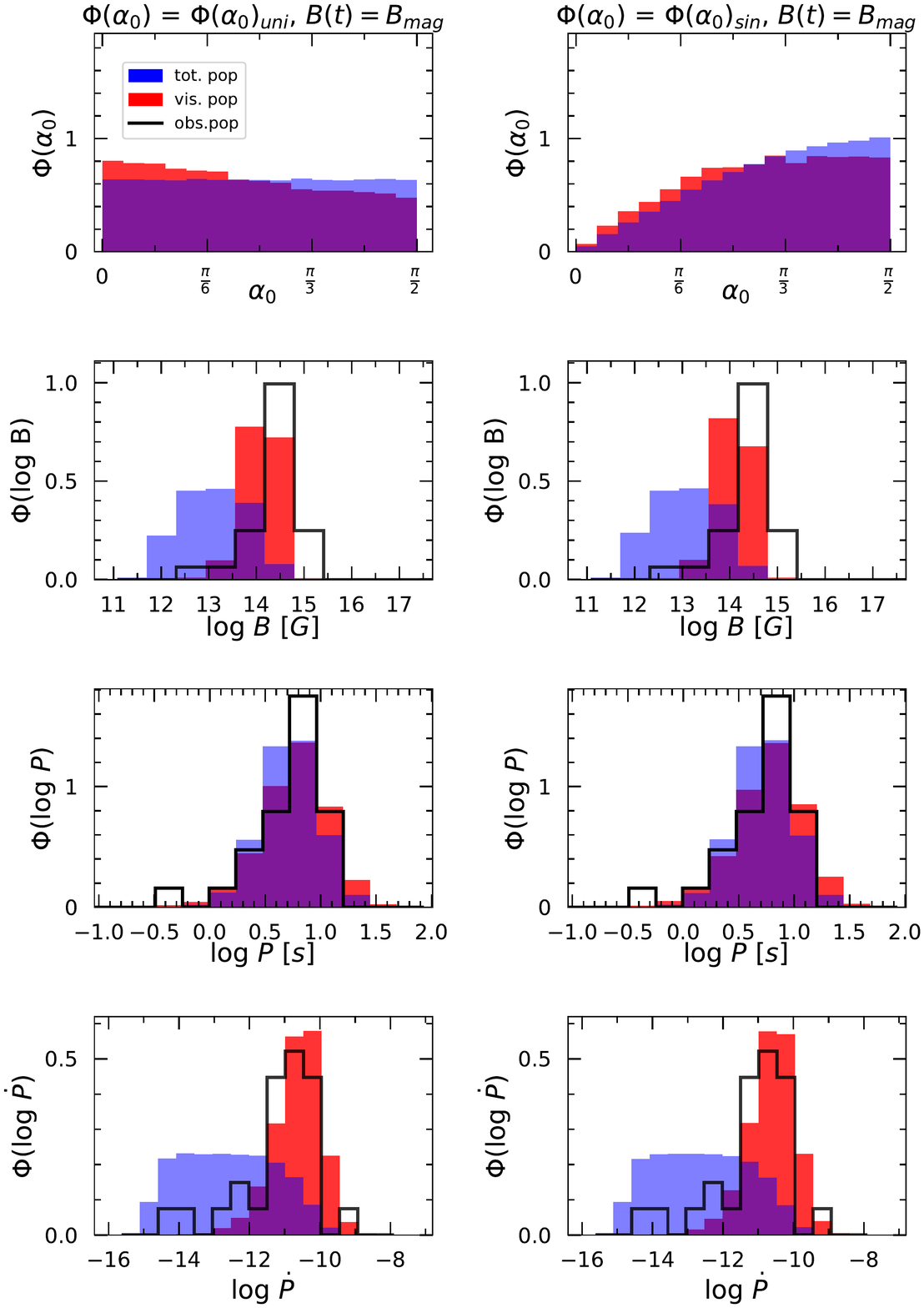}
\caption{Same as Fig.~\ref{fig:aB_hist1}, but for a synthetic population of magnetars for which $B(t)=B_{\rm mag}$.}
\label{fig:aB_hist2}
\end{figure*}

In order to determine how the different choices of $B(t)$ and $\Phi(\alpha_0)$ affect the visible populations of magnetars, we produce synthetic populations using the same $\vec{\theta}$ of input variables (equation~\ref{eq:theta-function}), but different $\Phi(\alpha_0)$ and $B(t)$ (Section~\ref{Chapter3_a_and_B}).
The values of $P$, $\dot P$, $\alpha_0$ and $B$ for both the total and visible populations are binned and normalized. Thus, eight probability density functions (PDFs) are produced for each population: $\Phi(P)_{\rm tot}$, $\Phi(\dot{P})_{\rm tot}$, $\Phi(B)_{\rm tot}$, $\Phi(\alpha_0)_{\rm tot}$, $\Phi(P)_{\rm vis}$, $\Phi(\dot{P})_{\rm vis}$, $\Phi(B)_{\rm vis}$ and $\Phi(\alpha_0)_{\rm vis}$ Here, the index {\em tot} refers to the total population (visible and faded magnetars) while {\em vis} refers to the visible population only. 
The ensemble of PDFs for the choice of $\vec{\theta}$ corresponding to $\vec{\theta}_{\rm opt}$ from model $A^M_3$ (see Table~\ref{tab:manualmodels}) are plotted in Figs.~\ref{fig:aB_hist1} and \ref{fig:aB_hist2}. 

The figures display (top to bottom panels) the normalized distributions, $\Phi$ of: initial magnetic inclination angles, $\alpha_0$; surface B-fields, $B(t)$; spin periods, $P$; and spin period derivatives, $\dot{P}$, for the total synthetic magnetar population (blue), the visible synthetic magnetar population (red), and the observed magnetar distribution (black contour). The left and right columns show the outcome of applying an initial distribution of magnetic angles, $\Phi (\alpha_0)$ that is uniform or weighted by $\sin \alpha_0$, respectively.
The free parameters used to produce the synthetic populations are the same as the ones from the optimized model $A^M_3$; that is: $\mu_{P_0} = 0.005\;{\rm s}$, $\sigma_{P_0}=0.5$, $\mu_{\dt{P}_0}=4.051\times 10^{-8}\;{\rm s\,s}^{-1}$, $\sigma_{\dt{P}_0}=1$, $s_1=3.610\times 10^{14}\;{\rm G}$, and $s_2=1.73$.
All synthetic populations contain 10\,000 visible magnetars and are evolved using avenue A, that is, with a constant magnetic inclination angle $\alpha = \alpha_0$.
Finally, the surface B-field distributions of the observed magnetars are produced by using the $P$ and $\dt{P}$ values from Table~\ref{table:ObsMag} and assuming an inclination angle of $\alpha (t)=\alpha_0=60^\circ$. There are no distributions of the observed magnetic inclination angles since they are unknown.

The difference between Figs.~\ref{fig:aB_hist1} and \ref{fig:aB_hist2} is the equation applied to estimate the surface B-field of the magnetars, $B(t)$. In Fig.~\ref{fig:aB_hist1}, we applied a pure vacuum dipole field ($B(t)=B_{\rm dip}$), whereas in Fig.~\ref{fig:aB_hist2} we applied the expression in equation~(\ref{eq:B(t)}) which combines a dipole field with a plasma-filled magnetosphere ($B(t)=B_{\rm mag}$). 

Certainly, the most significant differences are found between $\Phi(\alpha_0)_{\rm tot}$ and $\Phi(\alpha_0)_{\rm vis}$. No matter the choice of $B(t)$ and $\Phi(\alpha_0)$, $\Phi(\alpha_0)_{\rm vis}$ is shifted towards lower values compared to $\Phi(\alpha_0)_{\rm tot}$. This is most pronounced for the populations evolved using $B_{\rm dip}$ and $\Phi(\alpha_0)_{\rm uni}$.
Unfortunately, since the inclination angles of the observed magnetars are unknown, it is impossible to tell what choice is better.
It is also clear that applying $\Phi (\alpha_0)_{\rm uni}$ produces many more visible magnetars with $\alpha < \pi/6\;(30^\circ)$ compared to the case of applying $\Phi (\alpha_0)_{\rm sin}$.

The particular choice of $\vec{\theta}$ used here yields the best fit for the choice of $B(t)=B_{\rm dip}$ and $\Phi(\alpha_0)=\Phi_{\rm sin}(\alpha_0)$. This is obviously expected as model $A^M_3$ is optimized using this exact choice (see Section~\ref{sec:optimizing}). 
Performing this procedure on a couple of different choices of~$\vec{\theta}$, we find that the choices of $B(t)$ and $\Phi(\alpha_0)$ do not matter much. No matter the combination, a visible synthetic population that matches observations can be achieved as long as other parameters in $\vec{\theta}$ are adjusted accordingly.

\clearpage

\section{Manual models with zero-age and faded populations}\label{AppendixB}
Synthetic populations from all of the optimized models (see Tables~\ref{tab:manualmodels} and \ref{tab:automodels}) are plotted here. Both the zero-age and evolved magnetars are plotted, as triangles and circles respectively.  They are separated into the visible (red) and faded (gray) populations.

\begin{figure*}
\centering
\includegraphics[width=\linewidth]{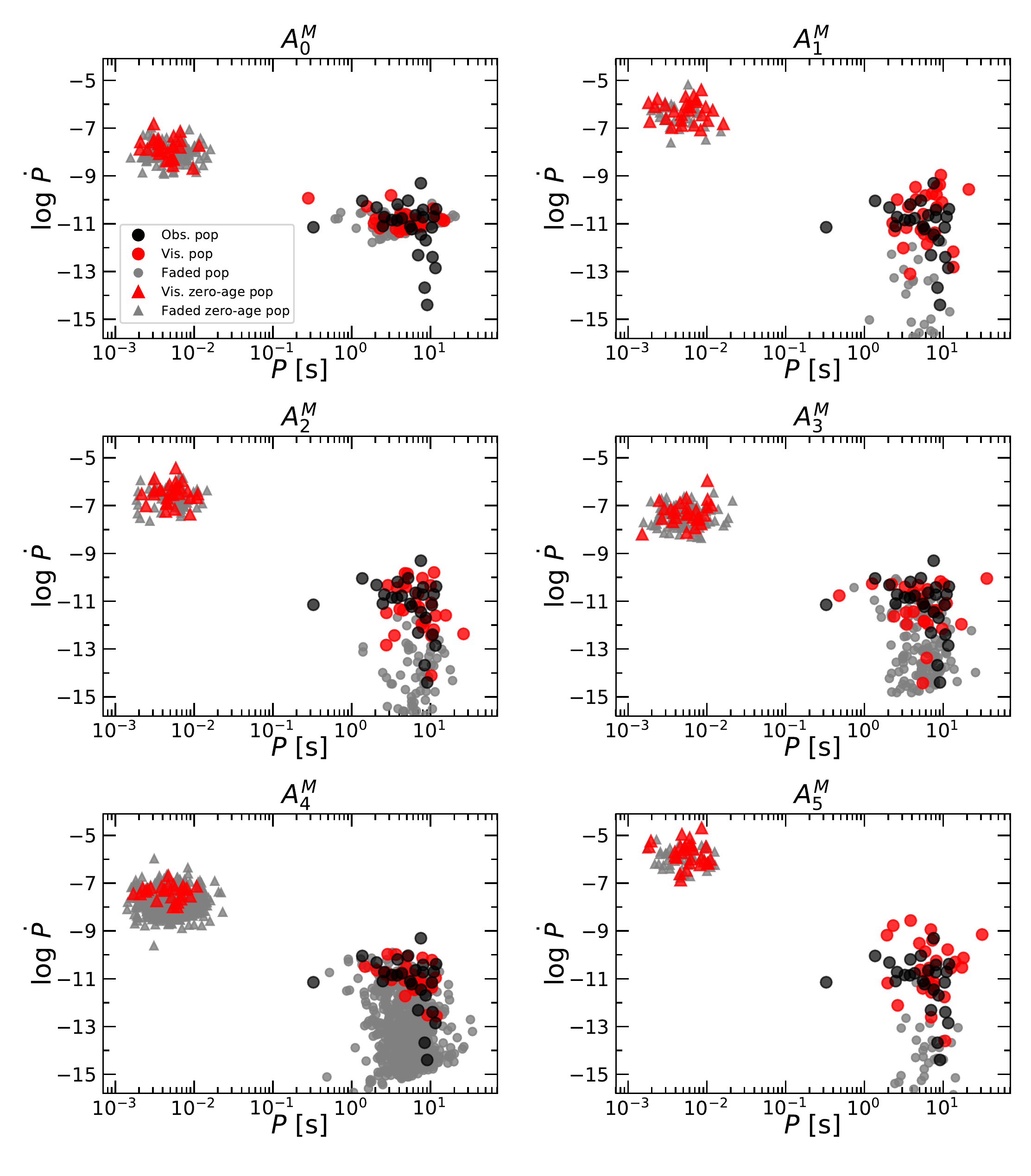}
\caption{$P\dot{P}$--diagrams. The synthetic populations are iterations of the models listed in Tables~\ref{tab:manualmodels} and~\ref{tab:automodels} produced using $\theta_{\rm opt}$ and $N_{\rm vis}=26$. The plots contain zero-age (triangles) and the current age (circles) locations for both the visible (red) and faded (grey) populations, together with the observed population (black).}
\label{fig:Full_plots_Appendix_1}
\end{figure*}

\begin{figure*}
\centering
\includegraphics[width=\linewidth]{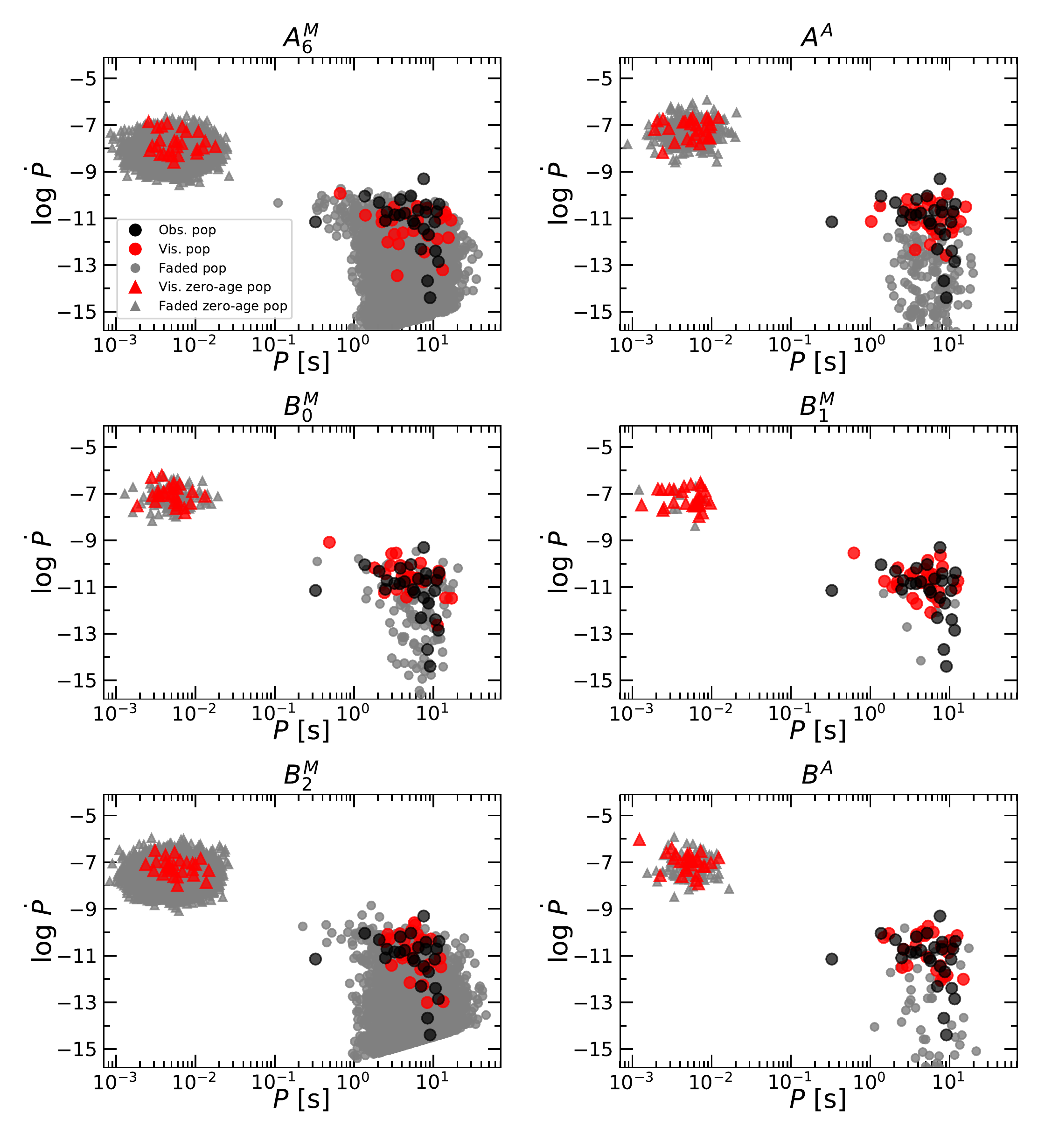}
\caption{Continuation of Fig.~\ref{fig:Full_plots_Appendix_1}.}
\label{fig:Full_plots_Appendix_2}
\end{figure*}

\clearpage

\section{Optimization Algorithms}\label{AppendixC}

\subsection{Manual algorithm}
The manual optimization algorithm is as follows:

\begin{enumerate}
\item Manually choose a value of all parameters in $\vec{\theta}$.
\item Generate a synthetic population with $N_{\rm vis}=100$ visible magnetars.
\item Calculate $S_{\rm vis}(P)$ and $S_{\rm vis}(\dot P)$ and perform two K-S tests to find the FOM.
\item Repeat steps (ii)$-$(iii) $N_{\rm ite}$ times and find the average FOM ($\mean{\rm{FOM}}$) and BR ($\mean{\rm{BR}}$).
\item Return to step (i) and choose a different $\vec{\theta}$.
\end{enumerate}

\begin{figure}
\centering
\includegraphics[width=\columnwidth]{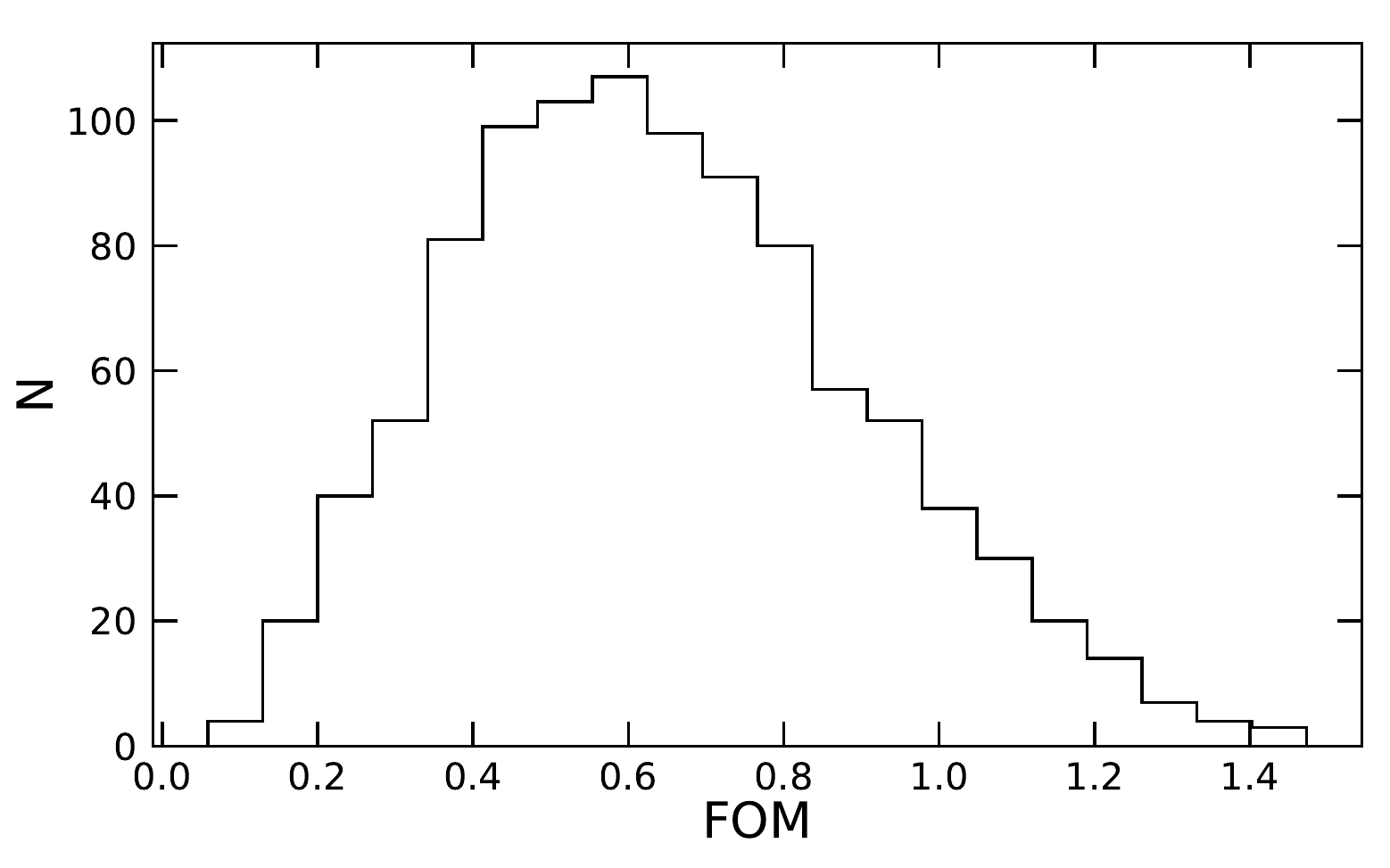}
\caption{Histogram of FOMs obtained from 1000 iterations of model $A_3^M$ (listed in Table~\ref{tab:manualmodels}). The average value is $\mean{\rm{FOM}}=0.653\pm0.016$.}
\label{fig:FOMhist}
\end{figure}

In the beginning, the number of iterations, $N_{\rm ite}$, is set to one. If the choice of $\vec{\theta}$ is very poor, then the synthetic population end up far away from the observed magnetars and one iteration is enough to tell if the fit is poor. As the synthetic populations start to lie closer to the observed one, $N_{\rm ite}$ is gradually increased to 1000. Such a high number of iterations is required in order to increase the accuracy of $\mean{\rm{FOM}}$. Due to the randomness of synthetic populations, the FOM can vary a lot. A histogram of 1000 FOMs obtained using the optimal $\vec{\theta}$ ($\vec{\theta}_{\rm opt}$) of model $A^M_3$ is plotted in Fig.~\ref{fig:FOMhist}.

This process is repeated by manually varying $\vec{\theta}$ until decreasing $\mean{\rm{FOM}}$ becomes impossible.
When this happens, we determine if the model is well fitting by checking if $\mean{\rm{FOM}} \la 1$ and $\mean{\rm{BR}} <25\;{\rm kyr}^{-1}$. If these hard constraints are met, we produce a parameter grid in the vicinity of the last solution and calculate $\mean{\rm{FOM}}$ for all points. This is done in order to confirm that the algorithm has converged. The $\vec{\theta}$ amongst the grid solutions yielding the smallest $\mean{\rm{FOM}}$ is then determined to be the optimal choice ($\vec{\theta}_{\rm opt}$).

No final grids were produced for models: $A_0^M$, $A_5^M$, $A_6^M$, and $B_2^M$, as they did not meet the constraints.

\subsection{Automatic algorithm}
The automatic algorithm performs a random walk through the space of free parameters. In the i'th step, new values for the parameters in $\vec{\theta}_{\rm i}$ are generated from log-uniform distributions. Afterwards, $\mean{{\rm FOM}}_{\rm i}$ is calculated together with the difference: $\Delta \mean{{\rm FOM}}_{\rm i} = \mean{{\rm FOM}}_{\rm i} - \mean{{\rm FOM}}_{\rm i-1}$.
If $\Delta \mean{{\rm FOM}}_{\rm i} < 0$, the step is accepted and saved. Otherwise, the algorithm uses a uniform distribution to generate a random variable, $R$, between 0 and 1, such that if:    
\begin{equation}
  R < \exp \left(-\frac{\Delta \mean{{\rm FOM}}_{\rm i}}{T} \right),
\end{equation}
then the step is accepted and saved despite the increase in $\mean{{\rm FOM}}$, otherwise it is rejected and the algorithm generates a new $\vec{\theta}$. This is repeated until a step is accepted or the limit of re-tries is reached.
Here, $T$ is the temperature parameter. The larger it is, the easier it is to accept a new step with $\mean{{\rm FOM}}_{\rm i} > \mean{{\rm FOM}}_{\rm i-1}$. In our case, $\Delta \mean{{\rm FOM}}_{\rm i} \in [0,2]$ facilitates $T \in [0.1,100]$. Near the upper limit of $T$, new steps always get accepted, while $T=0.1$ almost certainly requires $\Delta \mean{{\rm FOM}}_{\rm i} < 0$ in order to accept a new step.
By allowing for new steps to be accepted even though they have a larger $\mean{{\rm FOM}}$, it becomes possible for the algorithm to escape from local minima. 
$T$ is tuned such that any potential local minima can be escaped until the suspected global minimum is found \citep{gmvp14}.

This algorithm in repeated in cycles. One such cycle proceeds as follows:
\begin{enumerate}
\item Define the limits of all free parameters and choose $T$.
\item Generate $\vec{\theta}_i$.
\item Use $\vec{\theta}_i$ to generate and evolve a synthetic population with $N_{\rm vis}=100$. Repeat this $N_{\rm pop}$ times and find $\mean{{\rm FOM}}_{\rm i}$. If any population exceeds the BR limit of $20~\rm{kyr}^{-1}$, return to step (ii).
\item Calculate $\Delta \mean{{\rm FOM}}_{\rm i}$.
If $\Delta \mean{{\rm FOM}}_{\rm i} < 0$ or $\Delta \mean{{\rm FOM}}_{\rm i} > 0$ and $R < \exp \left(-\frac{\Delta \mean{{\rm FOM}}_{\rm i}}{T} \right)$ accept and save the new step and increment $i$ by 1. Otherwise, return to step~(ii).
\item Repeat steps~(ii)$-$(iv) until the desired number of accepted steps ($N_{\rm acc}$) is reached. 
\end{enumerate}

\begin{figure*}
\centering
\includegraphics[width=\textwidth]{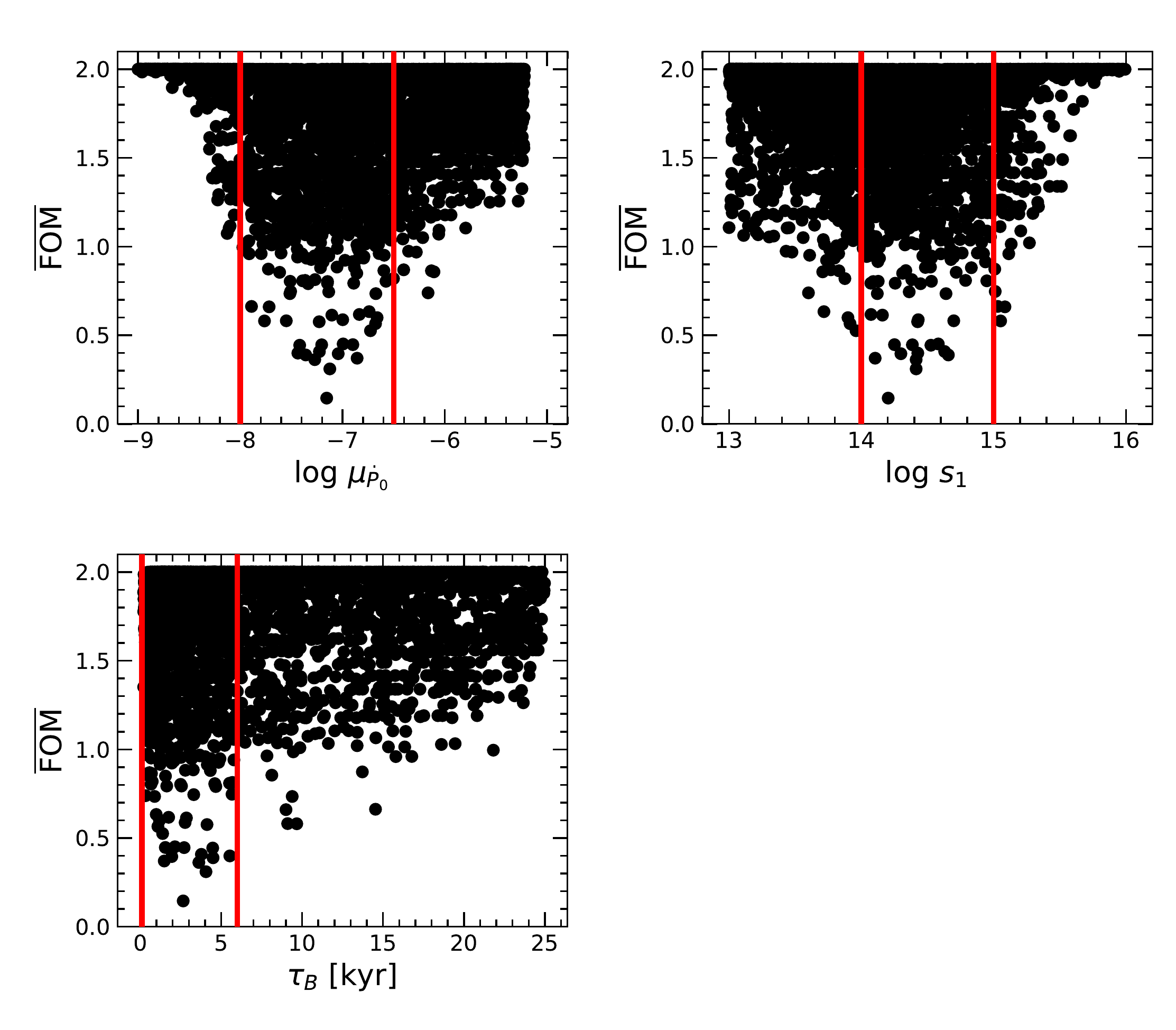}
\caption{Plots of $\mean{\text{FOM}}$ as function of free parameters for the first cycle of model $A^A$. The red lines mark the limits used in the second cycle.}
\label{fig:AppC_model_A_cycle_1}
\end{figure*}

At the end of a cycle, we plot $\mean{{\rm FOM}}$ as function of the free parameters. The resulting plots are used to set the limits for the next cycle. With each cycle, we also decrease $T$. In Fig.~\ref{fig:AppC_model_A_cycle_1}, the plots from the first cycle of model $A^A$ are plotted. In the figure, the limits chosen for the second cycle are marked by red lines.

In the initial cycles, we chose $N_{\rm pop}= 10$, $N_{\rm acc}=4000$ and $T=100$.
$N_{\rm pop}$ is gradually increased to 50, while $N_{\rm acc}$ is lowered to 1000 and $T$ to 0.1. In this way, the final cycles are more sensitive to small changes in $\mean{{\rm FOM}}$.
Of course, ideally, $N_{\rm acc}$ and $N_{\rm pop}$ would be kept large at all times, however we have to compromise due to limited computational resources.

\begin{figure*}
\centering
\includegraphics[width=\textwidth]{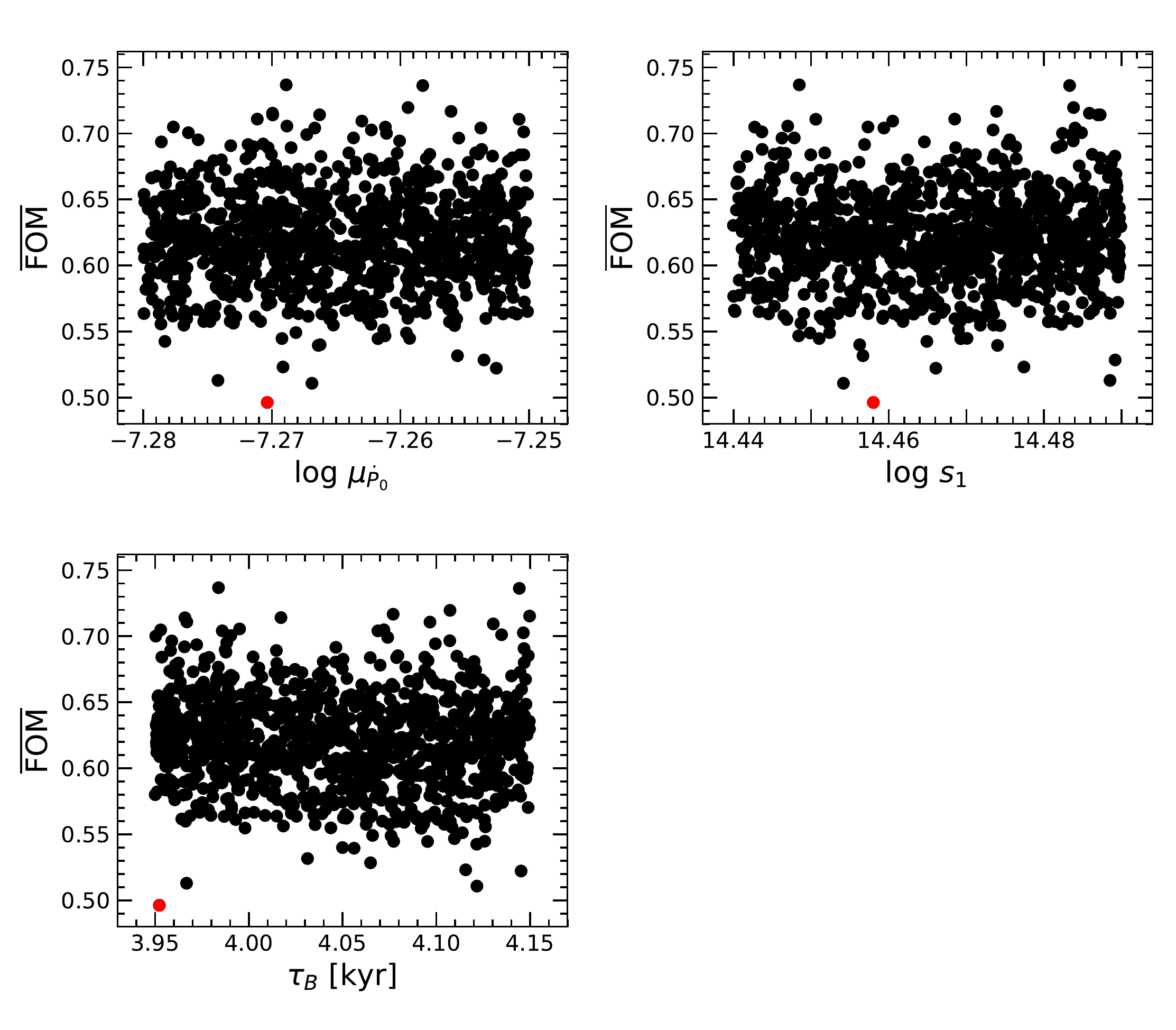}
\caption{Plots of $\mean{\text{FOM}}$ as function of free parameters for the last (16th) cycle of model $A^A$. The $\vec{\theta}$ that yields the lowest $\mean{\rm{FOM}}$ ($\vec{\theta}_{\rm opt}$) is plotted in red.}
\label{fig:AppC_model_A_cycle_last}
\end{figure*}

We stop iterating through new cycles when the plots of $\mean{{\rm FOM}}$ as function of the free parameters no longer exhibit any clear minimum. At this point, it is assumed that the algorithm has converged. The $\vec{\theta}$ with the lowest $\mean{{\rm FOM}}$ from the final cycle is defined as $\vec{\theta}_{\rm opt}$. The plots from the last (16'th) cycle of model $A^A$ are shown in Fig.~\ref{fig:AppC_model_A_cycle_last} with $\vec{\theta}_{\rm opt}$ being plotted as a red circle at the bottom. Unlike the plots from Fig.~\ref{fig:AppC_model_A_cycle_1}, there are no clear minima seen. However, this is expected for the narrow zoom-in of parameters in the 16'th cycle.

The more free parameters in $\vec{\theta}$, the longer it takes for the algorithm to converge. Due to this, $\mu_{P_0}$, $\sigma_{P_0}$, $\sigma_{\dot{P}_0}$ and $s_2$ are kept constant in all considered models.

\clearpage

\section{The two-sample Kolmogorov Smirnov test}\label{AppendixD}
For the purpose of the analysis, we use the two-sample K-S test which determines the likelihood that two empirical measures are drawn from the same underlying distribution, which does not have to be known.

The p-value of the K-S test is \citep{MarsagliaKS}:
\begin{equation}
  \text{p-value} = L(\sqrt{N} D)\;,
\label{eq:p-value}
\end{equation}
and takes on a value between 0 and 1.
Here $D$ is the D-statistic, defined as the absolute value of the largest difference between the ECDFs of the compared measures.
$N$ is the effective number of data points, $N = N_{\rm obs} N_{\rm vis} /(N_{\rm obs}+N_{\rm vis})$ \citep[]{PressKS}. Finally, L is the distribution function for $\sqrt{N} D$ \citep{MarsagliaKS}.

Naturally, under the null hypothesis (both the synthesized and the observed magnetar population are realizations of the same underlying population), the two ECDFs should be similar and the D-statistic is likely to be small, i.e. $D\ll 1$. Additionally, still
under the null hypothesis, the larger N is, the larger the likelihood of finding a small D-statistic.
The significance of the found D-statistic is quantified by determining the probability of it being equal to or larger than a chosen threshold value resulting from the null hypothesis.

In practice, the K-S test is used to find the optimal synthetic populations, i.e. meaning that the distributions of $P$ and $\dot{P}$ of the visible synthetic population resemble the ones of the observed magnetars as closely as possible.
This is accomplished by calculating the ECDFs of $P$ and $\dot{P}$ for both the visible population ($S_{\rm vis}(P)$, $S_{\rm vis}(\dot{P})$) and the observed population ($S_{\rm obs}(P)$, $S_{\rm obs}(\dot{P})$) and finding the D-statistics. Afterwards, two K-S tests are performed. One for $P$ and the other for $\dot{P}$. To accomplish this, we use the implementation of the two-sample K-S test from the SciPy Python module \citep{SciPy}.
The two p-values: $\text{p-value}(P)$ and $\text{p-value}(\dot{P})$ are used as measures of the goodness of fit. See Fig.~\ref{fig:KS_example1}.

A known issue of the K-S test is that it is not very sensitive to the tails of ECDFs \citep{RemyKS}. This can especially be a problem when working with small sample sizes, as is the case here. 
Another caveat is that the test ideally should be applied to measures of one-dimensional, independent variables (which is not the case for $P$ and $\dot{P}$, which are most likely dependent parameters). 

\clearpage


\section{Impact of varying $\eta$.}\label{AppendixE}

\begin{figure*}
\centering
\includegraphics[width = \textwidth]{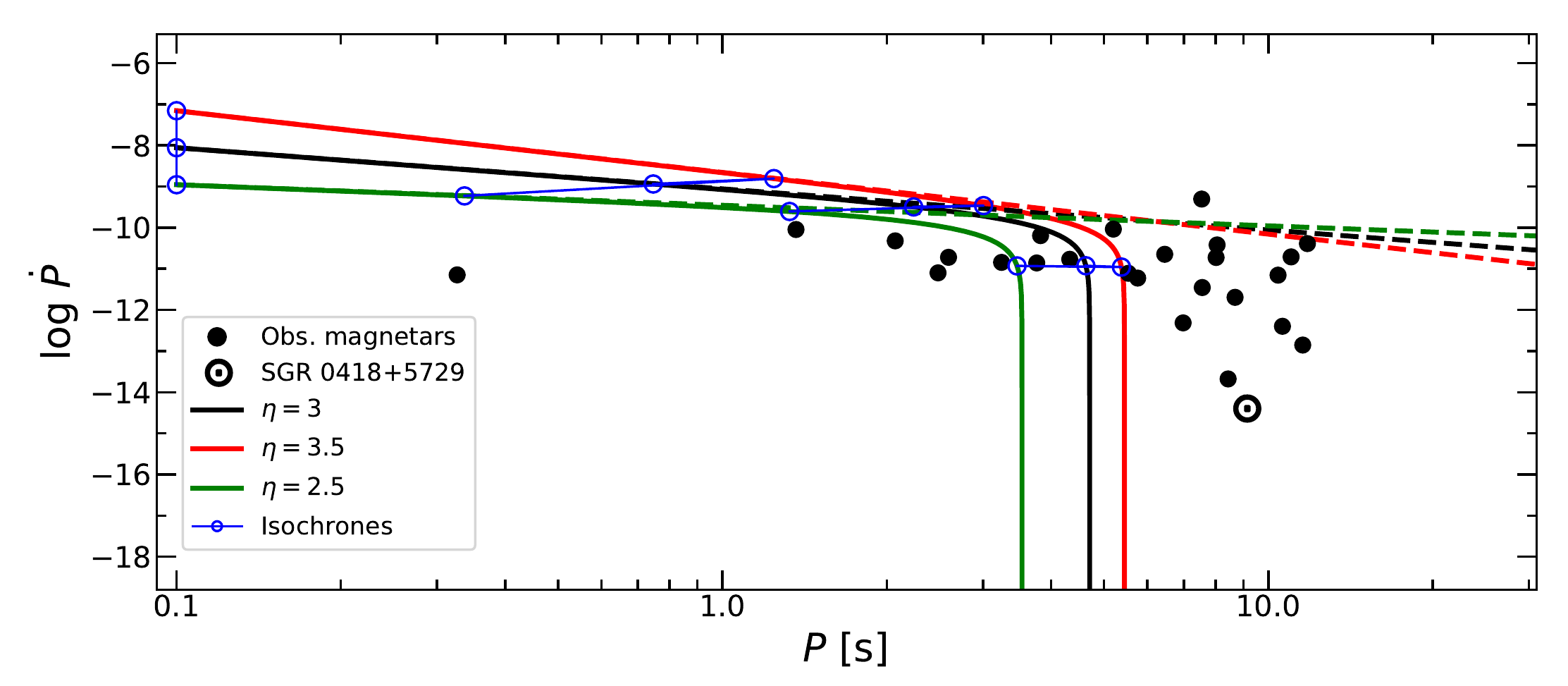}
\caption{Evolutionary tracks calculated using $\tau _B=1\;{\rm kyr}$, $\beta=-0.5$, $B_0 = 1\times 10^{15}~\rm{G}$, $P_0 = 0.1\;{\rm s}$ and different values of $\eta$. The solid black dots represent observed magnetars \citep{ok14}. The blue lines and circle are isochrones which mark the true ages, $t = \{0,\,0.01,\,0.1,\,1.0,\,10,\,100,\,1000\;{\rm kyr}\}$. The dashed lines are evolutionary tracks in the absence of B-field decay and alignment.  
SGR~0418+5729 \citep{rip+13} is shown with a circle.}
\label{fig:Evo_tracks_different_eta}
\end{figure*}
 
We have previously in Section~\ref{subsubsec:eta} argued for applying a constant value of $\eta =3$. Nevertheless, here we briefly discuss the cases for $\eta\ne3$.
Unlike $\beta$ and $\tau _B$, $\eta$ mainly influences evolution at $t<\tau_B$. Setting the initial value of $\eta>3$ causes $\dot{P}$ to decrease more sharply with $P$. Thus, increasing $\eta$ requires an increase of $\dot{P}_0$ in order for the synthetic magnetars to end up in the same region of the $P\dot{P}$--diagram. The opposite is true for $\eta<3$. Evolution tracks for three different choices of $\eta$ are plotted in Fig.~\ref{fig:Evo_tracks_different_eta}.

For a constant B-field, $\eta<2$ implies that $\dot{P}$ increases with increasing $P$. This is not always true when the B-field decays, as the decaying B-field counteracts this increase in $\dot{P}$.
In any case, we kept the initial value of $\eta$ between 1 and 4. Setting $\eta<1$ results in a negative exponent of the solution to equation~(\ref{eq:Dip_EroteqEdip_w_eta}). This has a large effect on how $P$ evolves with time and makes it hard to produce tracks that intersect with the observed magnetars.
Another reason for limiting this value has to do with the interpretation. Unlike $\beta$, which is easily interpreted as the parameter controlling the rate of B-field decay, the effects of $\eta$ are much more complex.

By setting $\eta \neq 3$, we essentially diverge from the dipole model. As long as the value is kept close to 3, the difference from the pure dipole scenario is not that large. In such a case, the model can be understood as a modified dipole model, perhaps something that takes into account the existence of a toroidal component, a multipole, or a plasma-filled magnetosphere.
The more $\eta$ diverges from 3, the harder it is to understand what physics the model in question actually represents.

\bsp	
\label{lastpage}
\end{document}